\documentclass[sigconf]{acmart}
\AtBeginDocument{%
  }

\setcopyright{acmlicensed}
\copyrightyear{2018}
\acmYear{2018}
\acmDOI{XXXXXXX.XXXXXXX}
\acmConference[Conference acronym 'XX]{Make sure to enter the correct
  conference title from your rights confirmation email}{June 03--05,
  2018}{Woodstock, NY}
\acmISBN{978-1-4503-XXXX-X/2018/06}

\usepackage{algorithm}
\usepackage[noend]{algorithmic}
\usepackage{xcolor}
\usepackage{microtype}
\usepackage{enumitem}
\usepackage{newfloat}
\usepackage{listings}
\usepackage{subcaption}

\usepackage{graphicx} 
\usepackage{courier} 
\usepackage{lipsum} 
\usepackage{amsmath}
\usepackage{natbib}

\usepackage{booktabs}
\usepackage{tabularx}
\usepackage{xurl}

\usepackage{tikz}

\newcommand{\system}[0]{\textsc{Tafo}}

\newtheorem{example}{Example}

\definecolor{jccolor}{rgb}{0.1,0.7,0.8}
\definecolor{vlcolor}{rgb}{0.9,0.1,0.1}
\definecolor{gcolor}{rgb}{0.7,0.3,0.7}
\definecolor{ccolor}{rgb}{0.3,0.3,0.7}
\definecolor{mrcolor}{RGB}{163,96,50}
\definecolor{mscolor}{RGB}{8, 102, 3}

\newcommand{\cststr}[1]{\text{\textcolor{gray}{"#1"}}}

\begin{document}

\title{\textsc{Tabularis Formatus}: Predictive Formatting for Tables}

\author{Mukul Singh}
\affiliation{%
  \institution{Microsoft}
  \city{Redmond}
  \country{USA}
}
\email{singhmukul@microsoft.com}

\author{Jos\'e Cambronero S{\'a}nchez}
\affiliation{%
  \institution{Microsoft}
  \city{Washington DC}
  \country{USA}
}
\email{josepablocam@gmail.com}

\author{Sumit Gulwani}
\affiliation{%
  \institution{Microsoft}
  \city{Redmond}
  \country{USA}
}
\email{sumitg@microsoft.com}

\author{Vu Le}
\affiliation{%
  \institution{Microsoft}
  \city{Redmond}
  \country{USA}
}
\email{levu@microsoft.com}

\author{Gust Verbruggen}
\affiliation{%
  \institution{Microsoft}
  \city{Keerbergen}
  \country{Belgium}
}
\email{gverbruggen@microsoft.com}

\renewcommand{\shortauthors}{Trovato et al.}

\begin{abstract}
Spreadsheet manipulation software are widely used for data management and analysis of tabular data, yet the creation of conditional formatting (CF) rules remains a complex task requiring technical knowledge and experience with specific platforms. 
In this paper we present \system{}, a neuro-symbolic approach to generating CF suggestions for tables, addressing common challenges such as user unawareness, difficulty in rule creation, and inadequate user interfaces.
\system{} takes inspiration from component based synthesis systems and extends them with semantic knowledge of language models and a diversity preserving rule ranking.
Unlike previous methods focused on structural formatting, \system{} uniquely incorporates value-based formatting, automatically learning both the rule trigger and the associated visual formatting properties for CF rules. By removing the dependency on user specification used by existing techniques in the form of formatted examples or natural language instruction, \system{} makes formatting completely predictive and automated for the user.
To evaluate \system{}, we use a corpus of 1.8 Million public workbooks with CF and manual formatting. We compare \system{} against a diverse set of symbolic and neural systems designed for or adapted for the task of table formatting. Our results show that \system{} generates more accurate, diverse and complete formatting suggestions than current systems and outperforms these by 15.6\%--26.5\% on matching user added ground truth rules in tables.
\end{abstract}


\keywords{Do, Not, Us, This, Code, Put, the, Correct, Terms, for,
  Your, Paper}


\maketitle

\section{Introduction}

\begin{figure}
    \centering
    \includegraphics[width=\columnwidth]{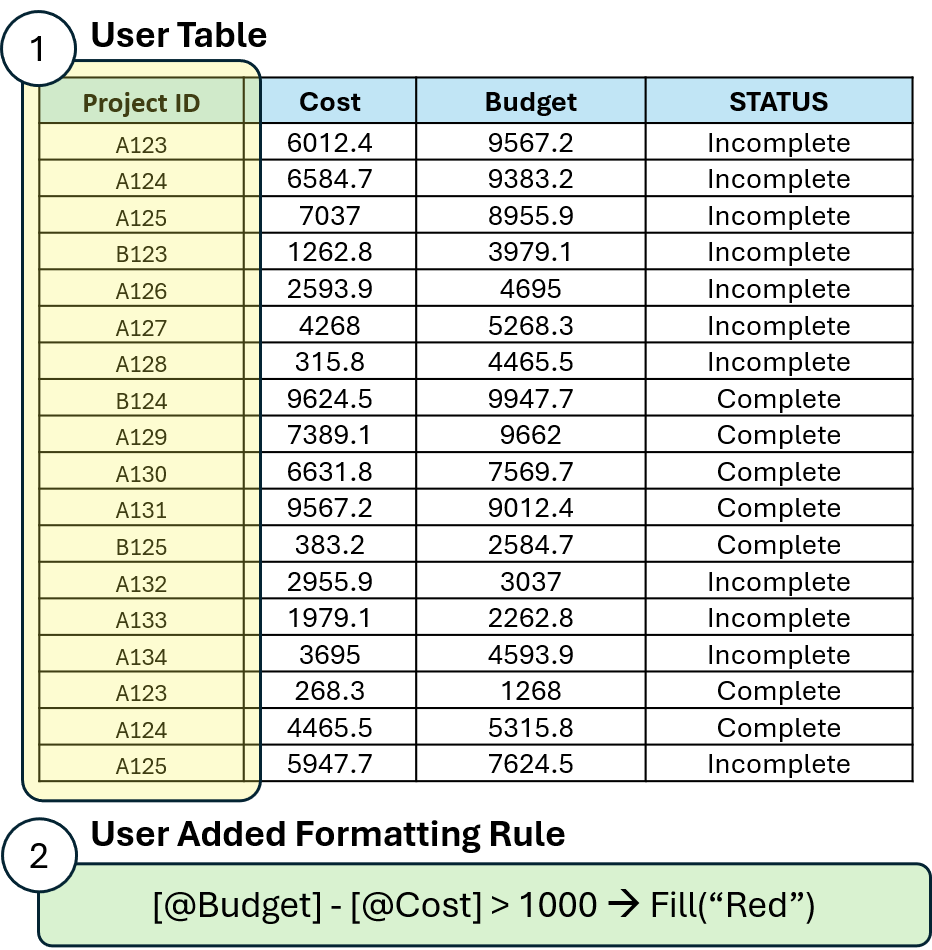}
    \caption{Sample data based formatting rule added by the user. (1) The user table in which the user wants to format the \textit{Project ID} column (highlighted in yellow). (2) The formatting rule added by the user to highlight \textit{Project ID} cells which are under budget (budget -- cost) by over 1000, in red color.}
    \label{fig:cf-example}
\end{figure}

Millions of users perform their data management and analysis in spreadsheet software \cite{spreadsheet-usage}.
To facilitate analysis and improve presentation, spreadsheet platforms typically allow users to write small, data-dependent formatting rules---also called \emph{conditional formatting (CF) rules}---that can help visualize their data and identify important takeaways. These rules are often based on data present in the table. Figure~\ref{fig:cf-example} shows an example CF rule added by a user to the table.

Previous work introduced the tasks of automatically learning conditional formatting rules from examples \cite{cornet} and natural language + examples \cite{format5}.
These systems while useful for automating tabular formatting. still impose two challenges for users: knowing \emph{what} to format---which requires understanding the data---and knowing \emph{how} to format it---which requires understanding the semantics associated with certain visual properties.
For example, the rule shown in Figure~\ref{fig:cf-example} applies a green fill color to Project ID cells where Budget is under Cost by over \textit{1000}.
In such a case, the green color fill clearly carries semantics (positive). 
Other cases, such as coloring losses or failure rates with red or making the maximum value in a column  bold, represent more common examples of formatting-based semantics.
Further, these systems rely on accurate user intents and \cite{format5} shows users struggle with providing complete and accurate specifications and the system has to ask for clarification.

In this paper, we go one step further in helping users with formatting their data.
We introduce \textsc{Tabularis Formatus} or \system{}, which predictively suggests conditional formatting rules without the need for communicating any intent.
Given only a table and a target column, \system{} automatically suggests relevant conditional formatting rules and the associated formatting properties.

Predicting conditions and their formatting properties raises three main challenges: (1) suggesting contextually relevant rules requires knowledge about the semantics of the data; (2) generated suggestions need to cover a diverse set of operations and executions (3) the learned rule formats need to be consistent with the data semantics and need to align with existing formatting in the table.

\system{} generates natural rules by combining a purely symbolic generator, a purely neural generator, and a neuro-symbolic  generator.
The purely symbolic generator enumerates rules by constructing predicates and combining them into rules with beam search and a trained heuristic.
The purely neural generator consists of an LLM Chain-of-Thought \cite{chain-of-thought} prompt, which incorporates basic table properties (like most common values and fraction of duplicates).
The neuro-symbolic learner leverages both symbolic and neural reasoning by extracting building blocks from the neural generator and using these in the predicate combination step with a higher weight.
To learn the formatting properties, \system{} mines a corpus for similar conditional formatting rules applied on columns and aligns them with the current sheet.

For training and evaluating \system{}, we use a corpus of 1.8 Million spreadsheet \cite{cornet} containing tables with CF rules and manual formatting. We found that \system{} can automate over 50\% of user formatting tasks for both CF and manually formatted tables with just 3 suggestions per task. We also compare \system{} to current automatic table formatting systems \cite{cornet, format5} as well as extend state-of-the-art language and table models to the task of CF suggestions. We find that \system{} consistently outperforms all baselines and has 15.6\%--26.5\% higher execution match accuracy on our benchmark.

In this work we make the following contributions:
\begin{enumerate}
    \item We introduce the task of predictive conditional formatting suggestions, consisting of both the rule condition and its formatting properties.
    \item We introduce \system{}, a neuro-symbolic system that provides conditional formatting suggestions.
    \item We evaluate \system{} extensively on data from 1.8 Million spreadsheets, comparing it to symbolic and neural baselines, and show that \system{} outperforms these by 
    \item We analyze the suggestions generated by \system{} for completeness, diversity, coverage and complexity.
\end{enumerate}
\section{Related Work}

Early work on rule based table formatting in spreadsheets has been relatively limited despite the large userbase of spreadsheet manipulation software. \cite{excelMathModel} explains the functionality and usage of conditional formatting in Excel. \cite{Abramovich04spreadsheetconditional} discusses rule based formatting and its application for demonstrating mathematical concepts.

Recent progress in CF has seen development of automated tools that learn CF rules based on user specification. The specification can be in the form of either examples \cite{cornet} or natural language instruction \cite{codet5+} or a mixture of both \cite{format5}. This line of work focuses on learning the CF rule specification through user supervision via multiple modalities. Unlike this, \system{} focuses on learning these rules purely predictively and does not require any supervision.

Another line of work on tabular formatting tackles automatic table formatting work which does not use data dependent rules. These include systems that rely on structural information from the table to learn formatting.
\cite{numerical_formatting} proposes CellGAN, a conditional Generative Adversarial Network model which learns the hierarchical headers and data groups in tables and can format these to visualize the data segmentation and improve presentation.
Similarly, \cite{rel_formatting_1,rel_formatting_2, payan2023instructexcelbenchmarknaturallanguage} focus on formatting cells based on table segmentation and cell classification.
These systems solely rely on the structure of the table and are targeted towards improving data presentation, they do not generate data dependent rules which are often essential to understand and bring out insights from the data.
In contrast, \system{} learns data dependent rules for the tables without supervision.

\system{} uses static data analysis to generate symbolic insights about the data which guide the component based synthesis and also assist the LLM in its reasoning. Such analysis has previously been used by \cite{auto-suggest} for learning data transformations. Techniques like this are also popular in other tabular tasks and have made their way into commercial spreadsheet software through systems like FlashFill \cite{flashfill} and FlashExtract \cite{flashextract}.
These systems, which are available in Excel, learn string transformation and data extraction programs using static analysis over string properties.

\system{} is a neurosymbolic system. It leverages language models to extract semantic knowledge relevant to the table to be used for symbolic CF synthesis. Previous neurosymbolic work have used LLMs as part of the generation pipeline \cite{format5}, for repair \cite{flame} for ranking \cite{llm-ranker}. All of these systems either use the symbolic reasoning to feed into the LLM or use the LLM to provide insights for the symbolic engine. Unlike these, \system{} is the first system where the symbolic reasoning is used by the LLM and the LLM reasoning is used for guiding the component based synthesis of rules.

Past work on generating predictive suggestions on databases have shown great success in the domain of querying \cite{vldb-query, vldb-search, vldb-pbe-1, datavinci} and data understanding and cleaning \cite{vldb-pbe-2}. \system{} builds upon these systems to solve the problem of table data formatting. In terms of ranking, previous work has explored program analysis \cite{ranking_program_feats, aligned-code-gen}, rule execution \cite{ranking_outputs}, diversity \cite{table-insights} or semantic relevance \cite{llm-ranker}. In this work, \system{} combines all of these to design a ranker that can generate relevant and diverse suggestions that are valid and executable.

Language models have revolutionized code generation \cite{codegen, codellama, codet5+, wizardcoder, starcoder, transcoder, codefusion, khatry2023wordscodeharnessingdata, vehicle-tele} and tabular analysis \cite{TAPAS, TaBERT, TabNet}. Specialized systems like \cite{spreadsheetCoder} and \cite{flame} are trained over millions of spreadsheets to generate formulas and other related artifacts. Models like TaBERT \cite{TaBERT} and TAPAS \cite{TAPAS} are popular Question Answering systems that use a neural model to encode the table and query. TUTA \cite{TUTA} is a weakly supervised model for cell and table type classification tasks.
Unlike these systems, \system{} uses a neurosymbolic technique that combines static data analysis and semantic information from language models for the task of learning formatting rules.
\section{Problem Definition}

Let $\mathbf{E} = [e_i^j]_{i = 1 \rightarrow n}^{j = 1 \rightarrow m}$ be a table with $n$ columns and $m$ rows. 
We will write $E_i$ and $E^j$ to denote column $i$ and row $j$, respectively. 
Each column $E_i$ is annotated with a column name (or header) $H_i$.
A cell $e$ is defined as a triplet $(v, t, \mathbf{f})$ of its value $v \in \mathcal{V}$, its annotated type $t \in \mathcal{T}$ and a mapping  $\mathbf{f}: F \rightarrow \mathcal{F}_F$ of formatting identifiers $F$ to one of their allowed values $\mathcal{F}_F$.
We call $\mathbf{f}$ the format of cell $e$, and write $v(e)$, $t(e)$ and $\mathbf{f}(e)$ to respectively denote its value, type and format.
We consider text, numeric and date as types, since these are the most common types used in spreadsheets \cite{cornet}.
We consider fill color, font color, bold, italics, and underline as formatting identifiers, as these are the most popular \cite{cornet-demo}.
$\mathbf{f}_\perp = \varnothing$ denotes no formatting.

\begin{example}
    In Figure~\ref{fig:cf-example}, the first cell in the ``Status'' column is represented as $e_4^1$ = (\cststr{Incomplete}, text, $\mathbf{f}_\perp$) as it is not formatted. We write $v(e_4^1) = \cststr{Incomplete}$.
\end{example}


We define a condition $C$ as a function $(\mathcal{V} \times \mathcal{T})^n \rightarrow \mathbb{B}$ that verifies some criterion on a row of $n$ cell values and types.
To ease notation in conditions, we will write $[@H_i]$ to refer to the implicit value or type of the cell in column $i$ of the given row.

\begin{example}
    In Figure~\ref{fig:cf-example}, the condition $$(v(R_3) - v(R_2))$$ is written as $$(\textsf{[@Budget]} - \textsf{[@Cost])}$$ because $H_2$ = \textsf{Cost} and $H_3$ = \textsf{Budget} and the \textsf{Subtract (--)} function is implicitly defined on the value of a cell.
\end{example}

A conditional formatting rule $R = (C, \mathbf{f})$ associates a condition with a format $\mathbf{f}$ to apply to cells that satisfy the condition in a target column $j$.
Given a table $\mathbf{E}$ and a target column $j$, the goal is then to suggest a list of \emph{desirable} conditional formatting rules.
Section~\ref{sec:metrics} describes the metrics that we use to measure whether suggestions are desirable with respect to ground truth conditional formatting rules mined from a large corpus of spreadsheets.




\section{Methodology} \label{sec:method}

\begin{figure*}
    \centering
    \includegraphics[width=\textwidth]{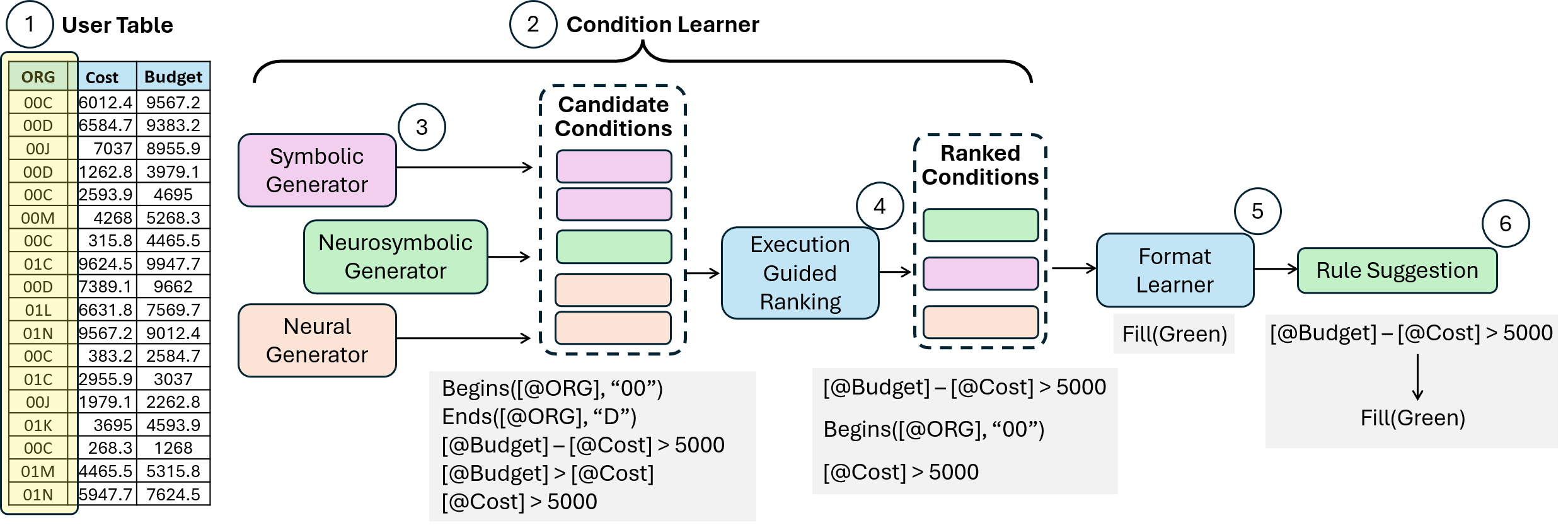}
    \caption{Summary of \system{} on a user table (1). First, \system{} learns multiple conditions for the tables (2) by pooling candidate conditions from multiple generators (3) and ranking them using an execution based ranker (4). After the conidition, \system{} generates the associated format for the learned conditionals (5) generating the final suggestion (6).}
    \label{fig:method-overview}
\end{figure*}

\system{} first learns multiple conditions $C$ from $(\mathbf{E}, j)$ by generating and ranking candidate conditions, and then learns the format $\textbf{f}$ from each $(\mathbf{E}, j, C)$.
We propose three different candidate generation strategies: purely symbolic, purely neural based on large language models, and a neuro-symbolic combination of them with bi-directional feedback.
An overview of this approach is shown in Figure~\ref{fig:method-overview}.
The following five sections respectively describe the three candidate generation strategies, ranking, and format learning.

\subsection{Learning conditions: symbolic} \label{sec:method-trigger}

\begin{figure}[htb]
    \centering
    \includegraphics[width=\columnwidth]{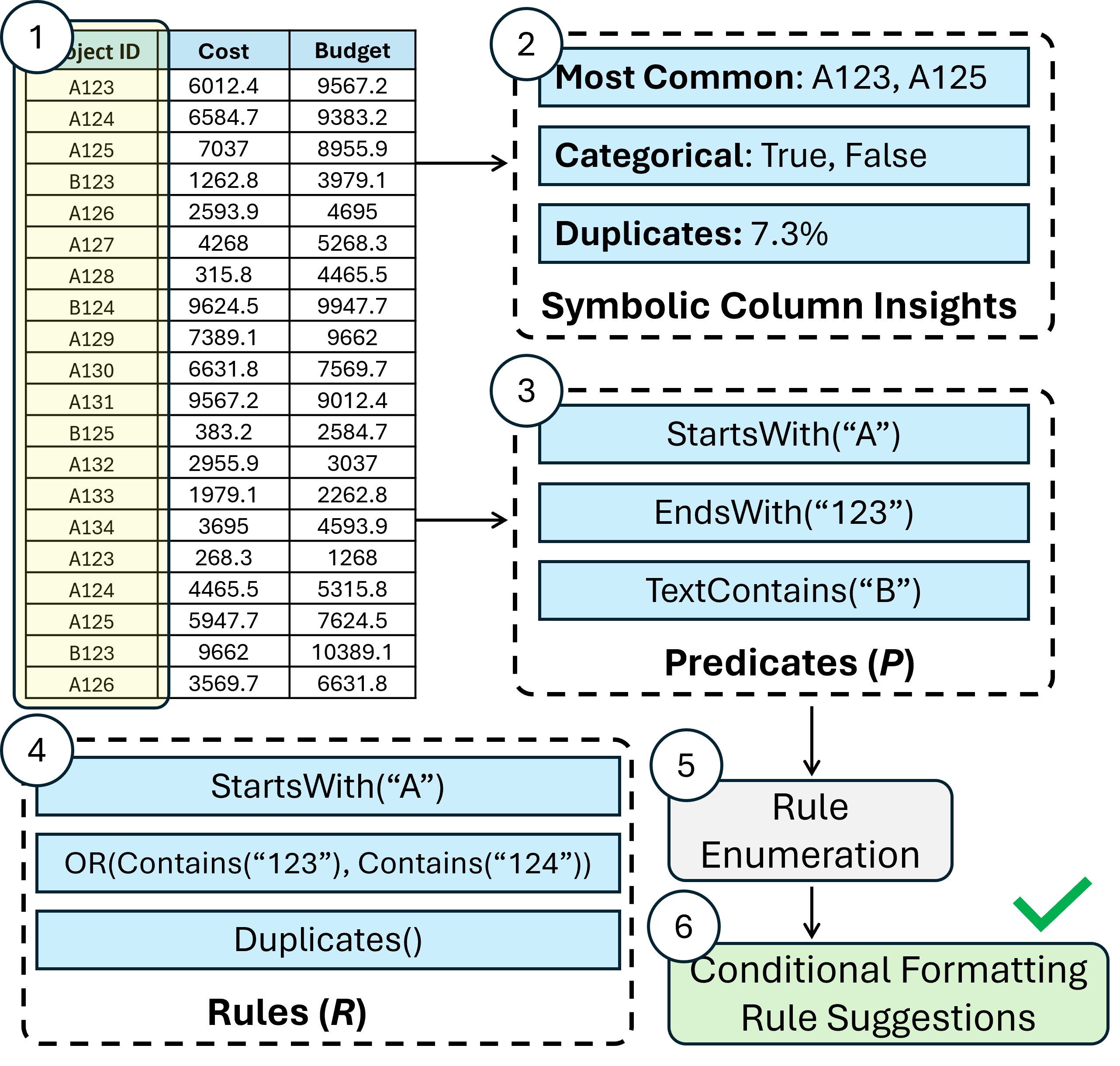}
    \caption{Symbolic rule synthesis overview. The input table (1) is used to extract table properties (2) that yield predicates (3) that are combined (4) into the final rules (5).}
    \label{fig:method-symbolic}
\end{figure}

On a high level, the symbolic generator first extracts signals in the form of static properties from the columns in the table, uses these properties to enumerate predicates, and then combines these predicates into complex conditions with beam search and a trained heuristic.
An overview is shown in Figure~\ref{fig:method-symbolic} and the following paragraphs describe the three main steps.

\paragraph{Extracting properties}

One way to generate properties is to symbolically construct them from base symbols.
We use a set of templates for this construction curated specific to formatting tasks, taking inspiration from other table tasks like cleaning \cite{datavinci}, transformation \cite{format5} and visualization \cite{table-viz}.
We compute properties over columns that are useful to reason about the formatting condition.
Table~\ref{tab:properties} summarizes the supported properties considered by \system{}.
Properties like \textsf{NumBlanks} and \textsf{NumFormulas} counting the occurrence of a specific condition.
Properties like \textsf{Average} and \textsf{Median} compute statistical properties.
Properties like \textsf{Categorical} and \textsf{FreeText} are heuristically computed to determine if the column contains categorical data and free text respectively.

One limitation of symbolically extracting properties is that the system has no semantic knowledge.
Because of this lack of semantic knowledge, it does not perform arithmetic columns to compute aggregate constants---training the ranker to understand which operations make sense based on only column names would require a \emph{huge} amount of data.
For example, in Figure~\ref{fig:method-overview}, the symbolic learner would never obtain \textsf{[@Budget] - [@Cost]} to reason about profits or losses.
LLMs, on the other hand, have been shown to capture much semantic knowledge about tables---they are increasingly used to suggest insights or transformations to users \cite{data-interpret,table-insights}.

\begin{figure}
    \centering
    \includegraphics[width=\columnwidth]{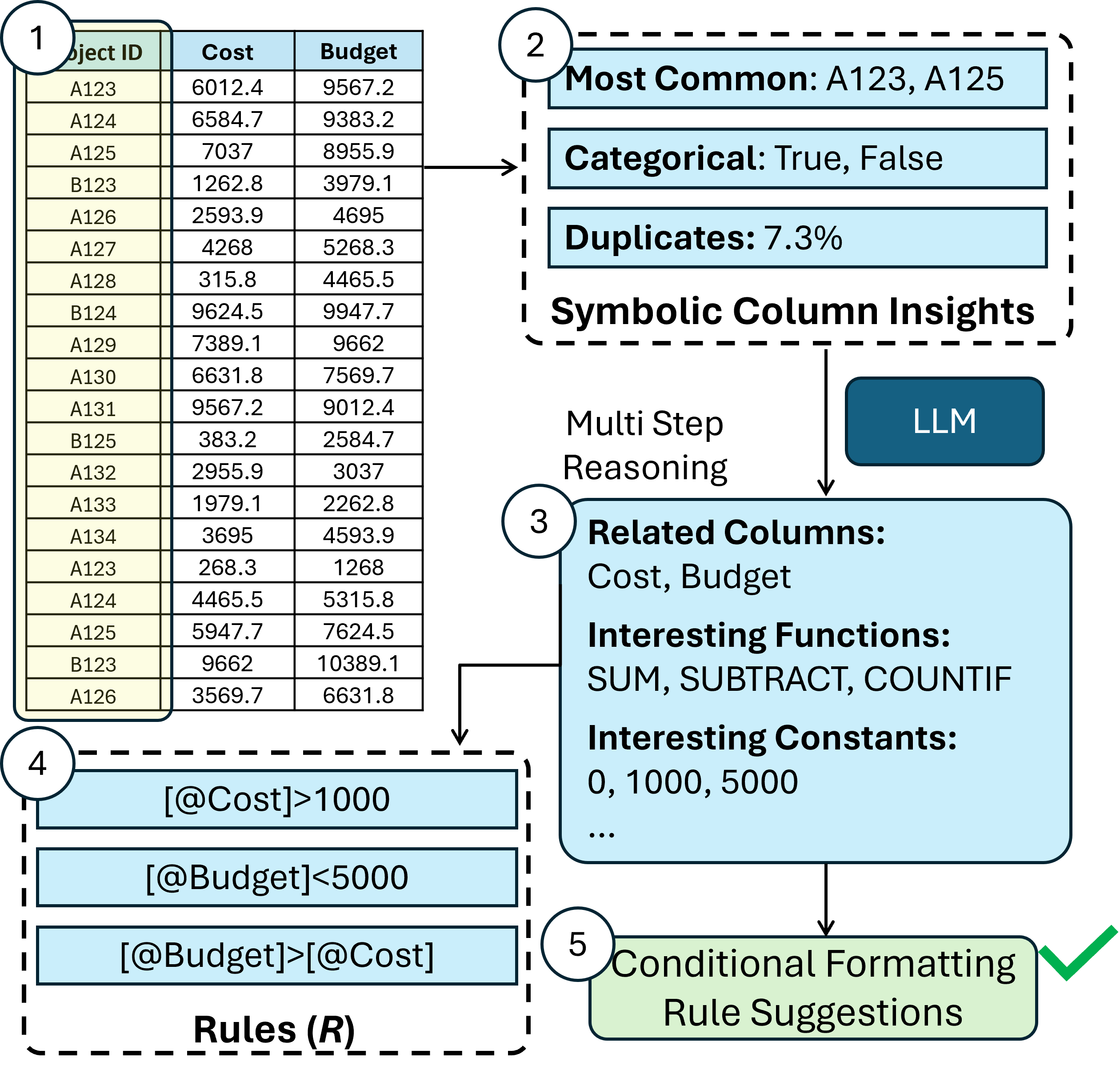}
    \caption{Overview of the LLM based Rule Generation. The input table (1) is used to extract table properties (2) which are used as context in the LLM prompt (3) followed by a multi step reasoning (4) and the final suggested rules (5).}
    \label{fig:method-neural}
\end{figure}

To leverage LLMs for generating conditions, we follow a multi-step reasoning approach, which is outlined in Figure~\ref{fig:method-neural}.
We use the column properties as table context in the prompt, along with headers, type info and sample values.
The model is instructed to generate the suggestions in four steps.
First, it identifies relevant columns it that should be formatted in the table.
Second, it generates useful predicates and functions that would be used in generating suggestions.
Third, it lists important constants that are relevant to the predicates and table.
Finally, using the previous steps, it generates a list of formatting rule suggestions.
To enforce the reasoning path, we provide three static examples in the prompt from the training corpus, which were manually annotated for the reasoning steps.
A sample of the prompt template is shown in Figure~\ref{fig:method-neural-prompt} and a sample of the full prompt is in the Appendix.

The generated rules are symbolically broken to generate base properties in the form of column names and constants and combined with properties and constants generated in the reasoning steps previously. This pooled set of properties is added back to the symbolically extracted properties.
Further, the neural generations also yields predicates (for example, \textsc{NOT(Blanks([@Column]))}) which are components constructed over properties (\textsc{Blanks}, \textsc{Column}) for better semantic knowledge distillation.


\begin{example}
    For the table shown in Figure~\ref{fig:method-overview}, the ``ORG '' column will have a text property, \textsf{MostCommonValues(3)} with value \{\cststr{00C}, \cststr{00D}, \cststr{00E}\} and a general property \textsf{NumBlanks} with value \textcolor{gray}{0}.
\end{example}

\begin{table}[htb]
    \small
    \centering
    \caption{Supported properties and their arguments for each datatype (top) and also general (bottom). The $k$ argument in properties determines the maximum number of items to be considered. For example, \textsf{Formulas(5)} has at most 5 formulas from the table. By default these are sorted by frequency.}
    \label{tab:properties}
    \begin{tabularx}{.9\columnwidth}{llX}\toprule
      Numeric     & Datetime  & Text  \\ \midrule
      \textsf{AverageValue} & \textsf{InLastWeek} & \textsf{MostCommonValues($k$)} \\
        \textsf{MedianValue} & \textsf{InNextWeek} & \textsf{DuplicatesValues($k$)} \\
        \textsf{90PercentileValue} & \textsf{InThisWeek} & \textsf{Categories} \\
        \textsf{75PercentileValue} & \textsf{InLastMonth} & \textsf{FreeText} \\
        \textsf{25PercentileValue} & \textsf{InNextMonth} & \textsf{AverageLength} \\
        \textsf{10PercentileValue} & \textsf{InThisMonth} &  \\
        \textsf{Skew} & \textsf{Today} &  \\
         & \textsf{Year} &  \\
      \midrule
      &General&\\ \midrule
      \textsf{NumErrors}& \textsf{NumBlanks} & \textsf{Formulas($k$)} \\
      \textsf{NumLogicals} & \textsf{NumNA} & \textsf{NumDuplicates} \\
      \textsf{NumUniques} & \textsf{NumDate} & \textsf{NumFormatted}\\  
    \bottomrule
    \end{tabularx}
\end{table}

\paragraph{Enumerating predicates} \label{sec:method-symbolic}

Predicates are boolean-valued function that takes a cell $e$ and with zero or more additional arguments, and return \textsf{true} if the property that it describes holds for the cell $e$.
All predicates are assigned a type and they only match cells of their type.
We use a synthesis based predicate generator that does bottom up construction over a collection of base properties and components to generate candidate predicates. Synthesis based predicate generators have shown good performance on formatting tasks \textsc{Cornet} \cite{cornet}. Unlike pervious work, we do two modifications that impact the quality of the enumerations -- (1) our synthesis is component based so the smallest element in the search space can be a combination of properties for example, \textsc{NOT(Blanks()} is a possible component; (2) the nodes in the sythesis process are ranked using a heuristic over the properties of the node.
For properties that return multiple values (like \textsf{MostCommonValues($k$)}) each output is considered as an individual constant.

\begin{figure*}
    \centering
    \includegraphics[width=0.95\textwidth]{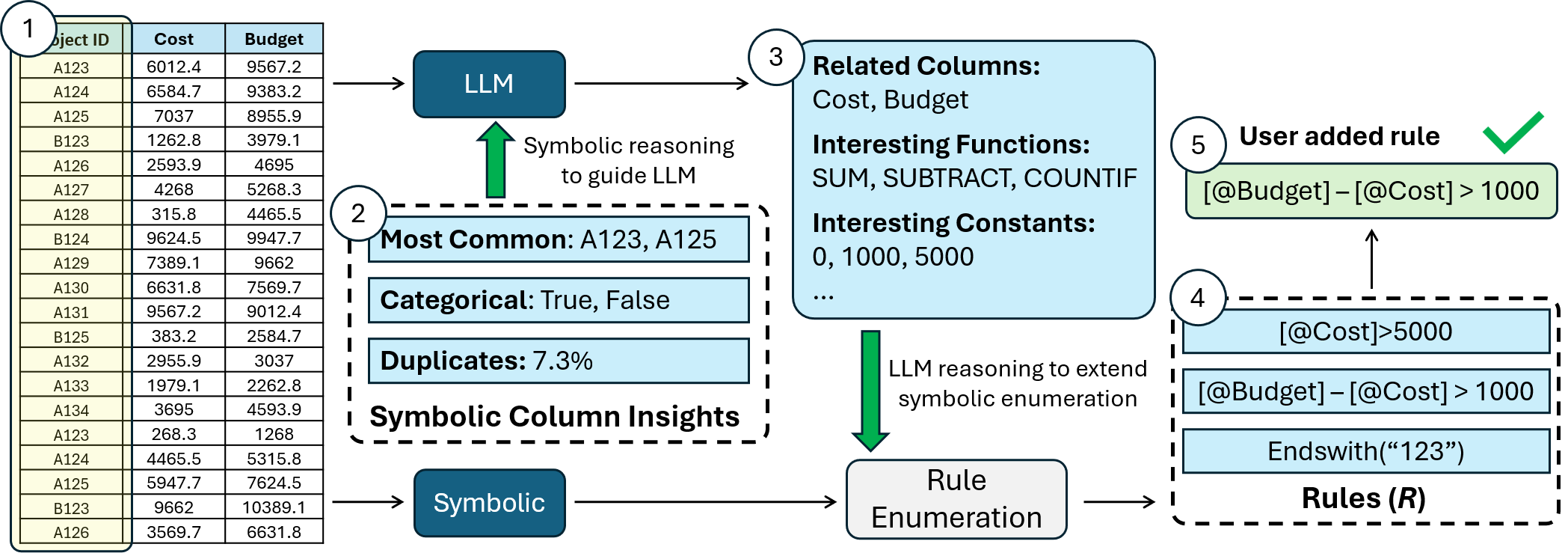}
    \caption{Figure summarizing the neuro-symbolic condition generator in \system{} for sample table (1). The symbolic column insights (2) are used as context by the LLM to generate conditions. The multi-step reasoning generated by LLM (3) is used by the symbolic rule enumerator to generate the final set of rules (4). This generates the user added condition (5) in the candidates.}
    \label{fig:method-neurosymbolic}
\end{figure*}

Neural properties and components are added to the rule enumeration step with higher weight, and constants and columns are used to boost the weight of symbolically generated predicates by adding a reward to the node scores which use these columns and constants.
The reward is computed as 10\% of the original score of the node.
Figure~\ref{fig:method-neurosymbolic} shows the overview of the neuro-symbolic generator on the example task.

\begin{example}
    For the ``ORG'' column in Figure~\ref{fig:method-overview}, two generated predicates are \textsf{TextStartsWith(\cststr{00C})} and \textsf{TextEndsWith(\cststr{D})}.
\end{example}

\paragraph{Combining predicates}

We represent complex conditions as propositional formulas in disjunctive normal form over predicates.
One challenge with enumerating rules is handling the huge number of candidates.
Prior work like \textsc{Cornet} \cite{cornet} and \textsc{FormaT5} \cite{format5} use an initial user specification---in the form of input-output examples and natural language descriptions, respectively---to guide this enumeration and prune the number of candidates.
We do not have any specification or user intent in our task.
Even with type constraints and syntax pruning, we end up with over 100K candidates per table.
We therefore train a ranker to score candidate rules---partial rules are valid rules---and use this ranker in a beam search to enumerate candidates.
We use the table properties (Table~\ref{tab:properties}) along with execution properties of the rule.
The properties of the partial program used for ranking are: percentage of table highlighted, type of rule, category of rule, and argument length.
We generate positive examples by mining partial rules from conditions in our corpus.
We generate negative examples by generating random rules over tables in our corpus.
The ranker is a dense network with three layers over the table and rule features, that is trained with logistic regression objective and used by considering the predicted logit as a score.



\begin{figure}
    \centering
    \includegraphics[width=0.9\columnwidth]{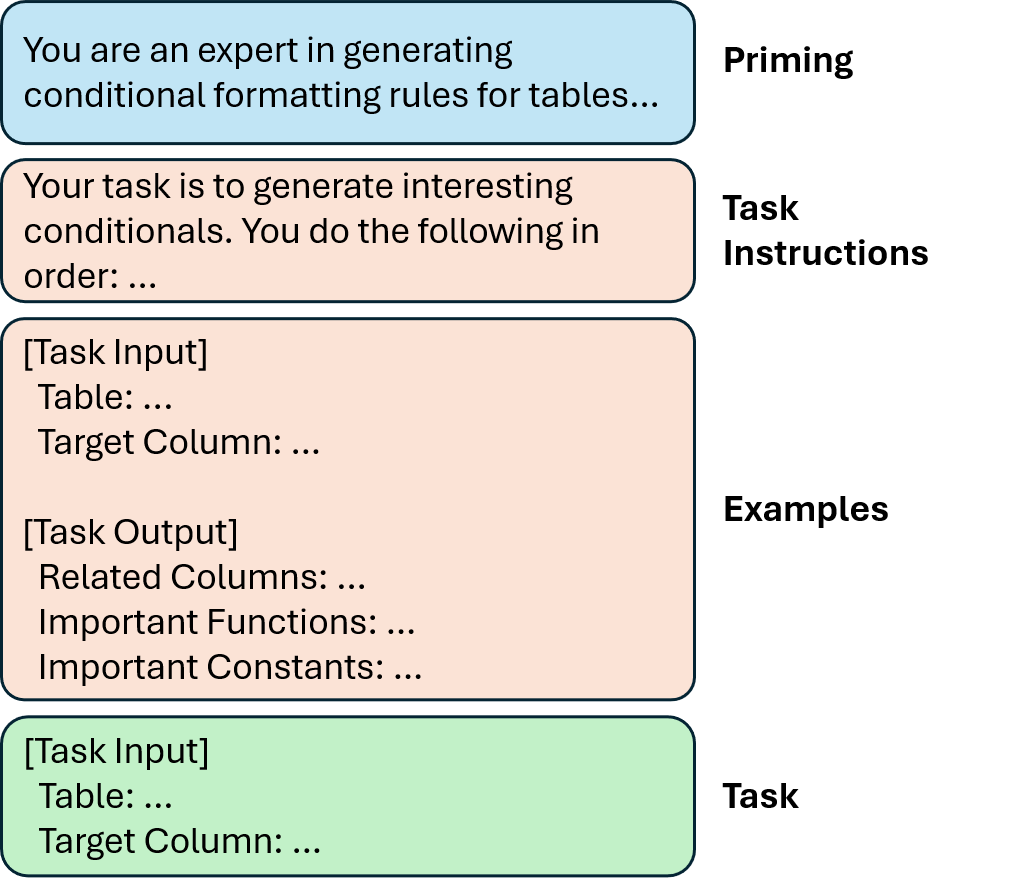}
    \caption{Prompt structure used to generate neural candidates for conditionals. The prompt uses a multi-step reasoning approach in a few-shot setting with task priming.}
    \label{fig:method-neural-prompt}
\end{figure}

\subsection{Why we need neurosymbolic generation} \label{sec:method-neurosymbolic}

We can consider a pure symbolic systems that extracts properties using templates and does rule based enumeration for predicate construction, or a pure neural system that prompts a language model to directly to generate predicates as alternatives to the neurosymbolic learner.
Upon examining the candidates generated by pure symbolic and neural systems---see Figure~\ref{fig:method-symbolic} and Figure~\ref{fig:method-neural}---we find that none of these generate the actual user added condition, \textsf{[@Budget]-[@Cost]>1000}.
Symbolic candidates contain diverse operations, but they lack semantic knowledge to know that ``Budget'' and ``Cost'' are related.
Neural candidates have semantic knowledge, but they lack the diversity due to lack of execution.
For example, the top neural candidate \textsf{[@Budget]>{@Cost}} evaluates to true for every row in the table and upon execution formats the entire column.



Our experiments show that considering the neural, symbolic and neuro-symbolic approaches separately and then ranking still improves the performance over only using the neuro-symbolic approach (Section~\ref{ssec:design}).
This happens due to the increased diversity, which significantly impacts performance (Section~\ref{ssec:exp-analysis}).


\subsection{Ranking conditions} \label{sec:method-ranking}

\begin{figure}
    \centering
    \includegraphics[width=\columnwidth]{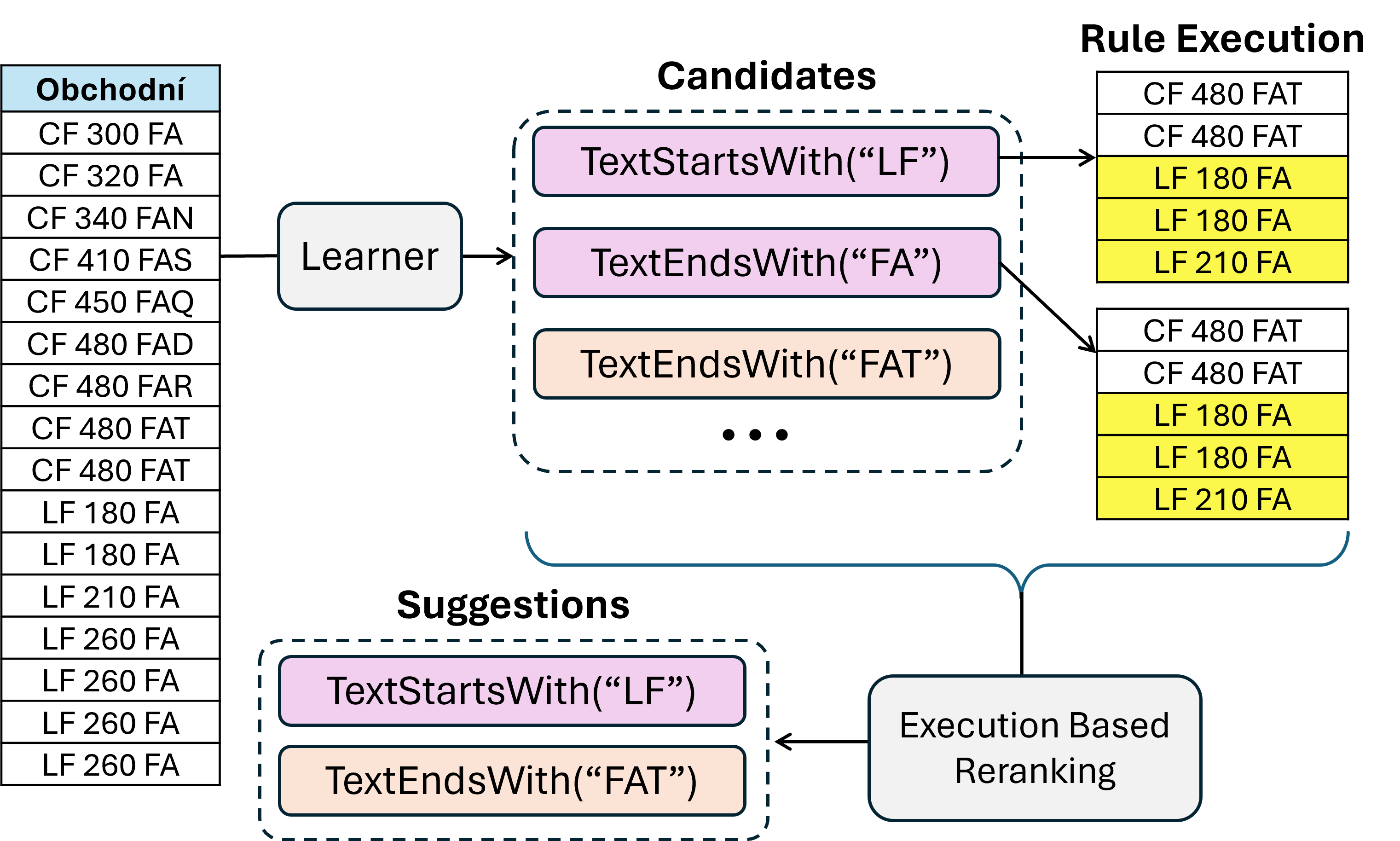}
    \caption{Suggestion ranking system. The learner generates a ranked set of candidate rules based on the applicability of the rule. To ensure a diverse selection of suggestions shown to the user, \system{} executes the candidates rules and clusters the candidates into equivalence classes based on the output formatting they generate. The clusters are then ranked with the top ranked rule from each cluster being selected.}
    \label{fig:method-ranking}
\end{figure}

The generated conditions need to be ranked based on their relevance to the table.
Previous work has used symbolic features \cite{ranking_program_feats}, LLMs \cite{llm-ranker} and custom models \cite{cornet} for ranking suggestions.
One key difference in formatting is that its a visual task---the execution of the rules play a crucial role in determining its score.
Furthermore, we also want the generated suggestions to be diverse and cover different parts of the rule grammar and the table.

To solve these competing objectives of diversity and quality, we use execution-based ranking.
Let $\mathbf{C}$ be the collection of learned conditions.
First, we use the node ranker described in Section~\ref{sec:method-symbolic} to assign a score $s_i$ for all conditions $C_i \in \mathbf{C}$.
Next, we execute all candidate rules $C_i \in \mathbf{C}$ on the table $\mathbf{E}$ to get the output as boolean vectors $O_i \in \mathbb{B}^n$.
The candidates are then clustered into semantic equivalence classes $EC_k$ such that $\forall C_i, C_j \in EC_k : O_i = O_j$.
Each equivalence class is a unique execution on the column.


Each cluster is scored based on the average scores of the rules present in the cluster $S_i = Avg(s_r \in EC_i)$. 
To generate the final list of rules we do a round robin sampling of the rules in order of the rule scores $s_r$ from clusters in order of the cluster scores $S_i$ for all clusters $\bigcup_{n} EC_i$. This ensures that the rules belonging to different execution equivalence classes are prioritized.
Figure~\ref{fig:method-ranking} summarizes the overview of the execution guided ranking setup.
\begin{example}
    Consider the table shown in Figure~\ref{fig:method-ranking}, the learner generates candidates which are then clustered into semantically equivalent classes denoted by orange and pink highlighting. The top rule from each semantic class is picked as the final ranked subset.
\end{example}

\subsection{Learning formats} \label{sec:method-format}

\begin{figure}
    \centering
    \includegraphics[width=\columnwidth]{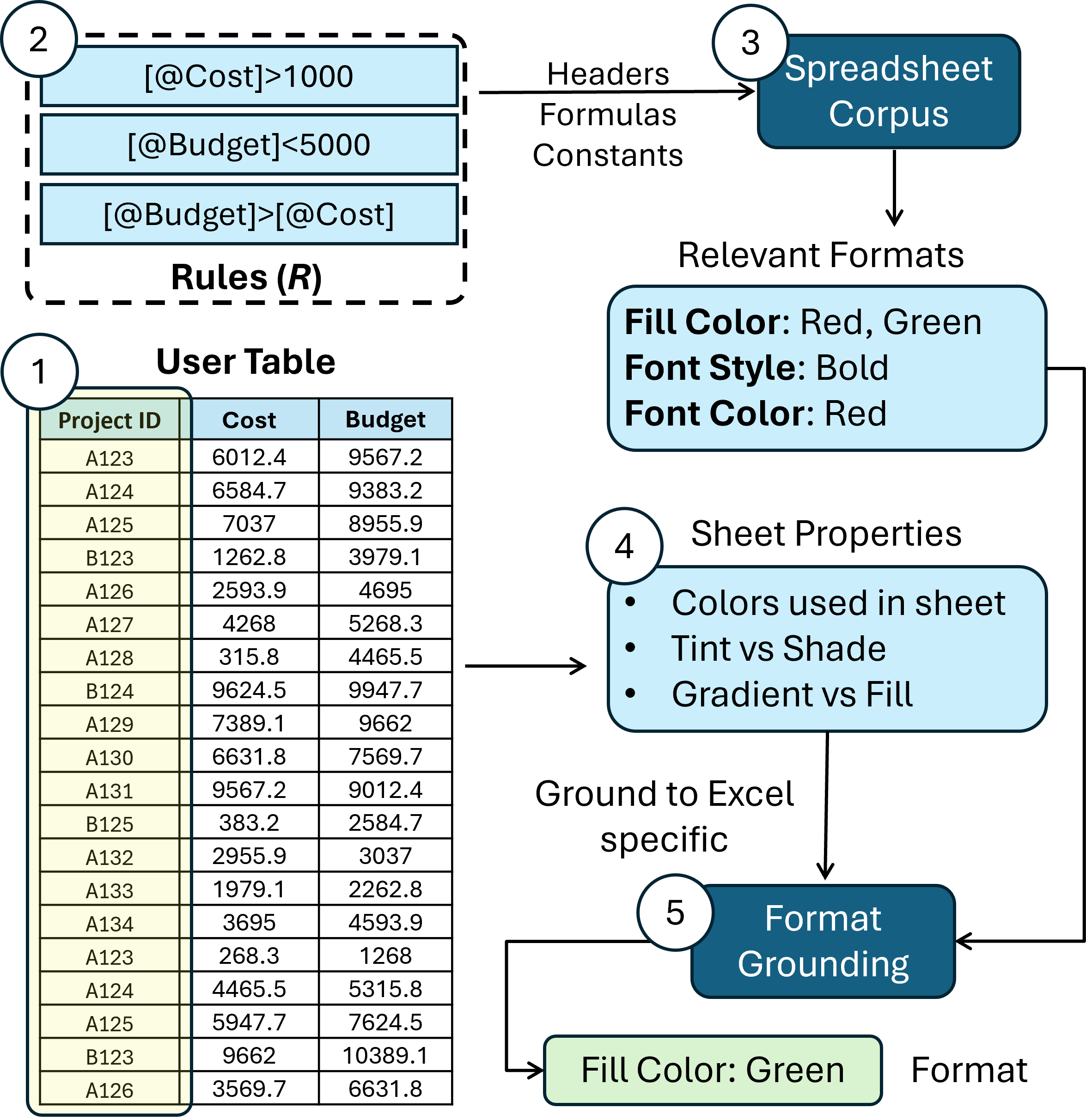}
    \caption{Summary of the format learning for \system{}. Given a user table $\mathcal{T}$ (1) and the learned rule triggers $t \in (t,f) = r \in R$, \system{} retrieves similar $(table, rule)$ pairs from the corpus (3) and mines formats used in them. The mined formats are then transformed based on the formatting properties found in $\mathcal{T}$. The final formats are grounded and the final format is returned (5)}
    \label{fig:method-format}
\end{figure}

Since conditional formatting is a visual task, suggesting the right format is almost as crucial as learning the condition.
For example, a condition \textsf{TextEquals(\cststr{PASS})} is more likely to be associated with a light green fill color than with a red font color.
Past work \cite{cornet, format5} has not looked at learning formats and only focused on learning the rule component of formatting rules.
We build on the intuition shown by previous work in spreadsheet automation \cite{nl2formula} which suggest that similar tables have similar artifacts, like formulas, charts and pivot tables.
We extend this hypothesis for formats used in tables.
We leverage a corpus of conditional formatting rules to suggest relevant formats for a given condition $C$ and table $\mathbf{E}$.

First, we collect candidate formats $\mathbf{f}_c = \mathbf{f}_e \cup \mathbf{f}_r$ with $\mathbf{f}_e$ being all formats already present in the current sheet and $\mathbf{f}_r$ the formats from similar tables and rules in the corpus.
To find these related tables, we compare tables on their headers and formulas they contain, and rules on the predicates and their constants of the condition $C$.
We use weighted hamming distance to compute this similarity, where weights are obtained by manually annotating the similarity of 100 tables and performing linear regression over these.
We mine rule-table pairs in decreasing order of their similarity until either $\lambda_N$ pairs have been mined or the similarity score drops below a threshold $\lambda_T$, with $\lambda_N$ and $\lambda_T$ hyperparameters.


Second, we select the formats using statistical properties of the format identifiers and their values.
We score format identifiers based on their frequency in the candidate set.
To bias the system towards formats used in the current sheet, formats in $\mathbf{f}_e$ have twice the weight as those in $\mathbf{f}_r$.
We pick all format identifiers that have an average score greater than 0.5 in the candidate set.
To pick values for the format identifiers, we use the same frequency based ranking withing the format identifier with a slight change that colors are compared using the closest HTML web color instead of direct hex codes.
For each format identifier, we only keep the highest ranked value.
The final format is then the union of all selected formats.

\begin{figure}
    \centering
    \includegraphics[width=\columnwidth]{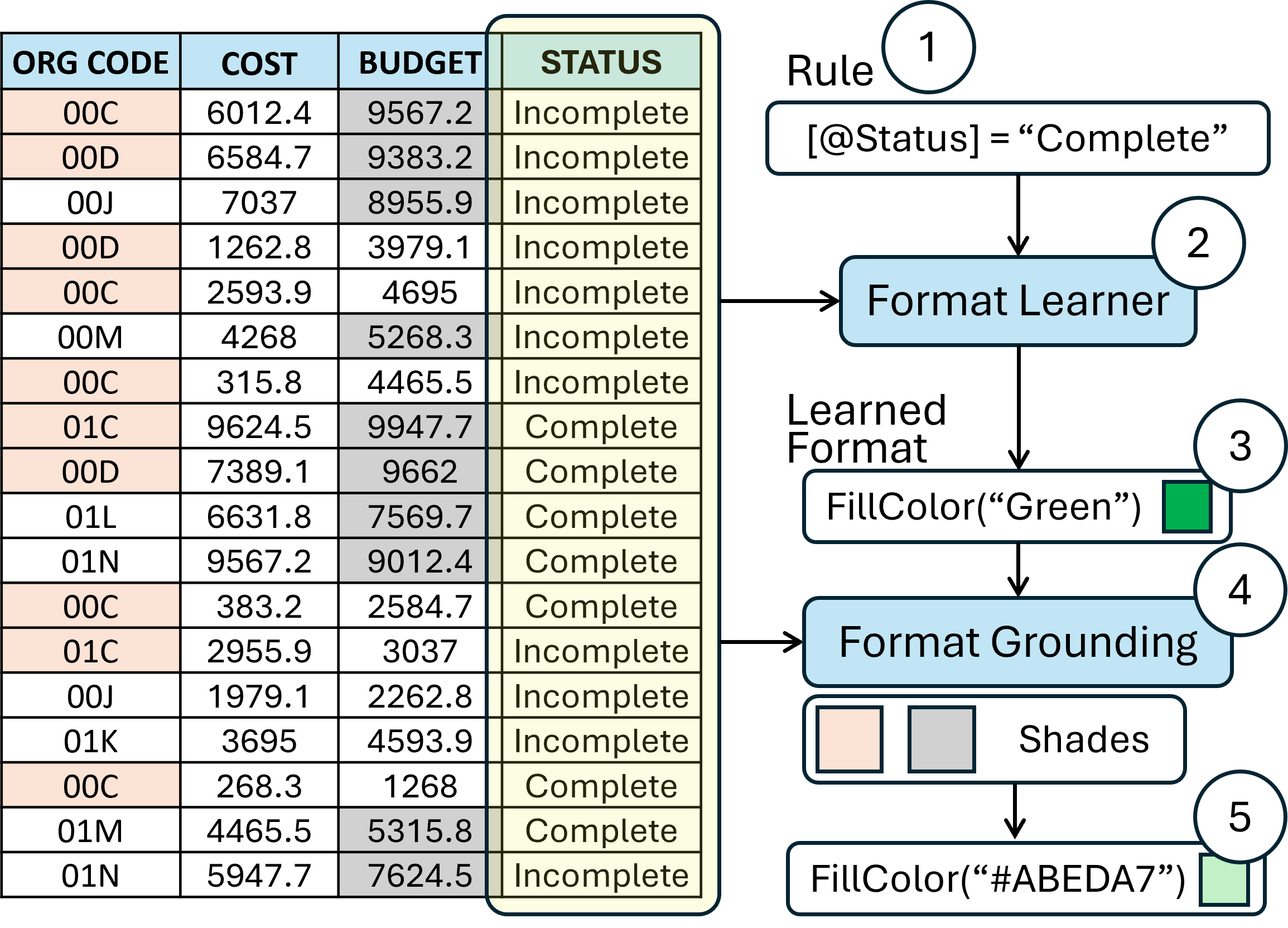}
    \caption{Figure showing the format grounding for a sample table: (1) the learned rule for which we need to learn formats; (2) \system{} format learner using corpus and table information; (3) the formats learned; (4) The table has all existing formatted cells in shade colors and not solid colors; (5) The suggested format is a shade version of the originally learned format.}
    \label{fig:method-format-grounding}
\end{figure}

Finally, since Excel allows using the same color with different shades, we ground the format properties based on indicators mined from the table.
If at least 75\% of cells has a specific shade, we suggest that shade.
This allows \system{} to personalize the formats to user preferences beyond specific colors.

\begin{example}
    Figure~\ref{fig:method-format-grounding} shows a sample table where the rule to highlight cells equal to \cststr{Incomplete}. The learned format for this rule is \textsf{FillColor(\cststr{\#008000})} which is solid green. The format grounding engine mines the shade indicator since all existing formats in the table have shade fill colors. The format grounding thus grounds the fill color to \cststr{\#ABEDA7} which is a lighter shade of solid green.
\end{example}

\section{Evaluation Setup}
In this section we describe the benchmarks, evaluation metrics and baseline systems which we use for evaluating \system{}.

\subsection{Benchmarks} \label{sec:benchmarks}

To train and evaluate \system{}, we leverage a corpus of 1.8 million Excel workbooks scraped from public sources \cite{cornet}.
This corpus has previously been used to study tabular formatting tasks~\cite{cornet, format5}.
Each task in this corpus consists of a table and a conditional formatting rule.
We use the de-duplicated set as described in \cite{cornet}, which consists of 105K tasks.
We use 80K tasks for training the ranker and rule generation engine,
and the remaining 25K tasks for evaluation.
Table~\ref{tab:benchmark-tasks} shows a summary of the tasks (1).

Additionally, we also consider non rule based table formatting. We extract tables with columns that contain manual formatting for cell fill color (without conditional formatting rules).
We consider a column $E^j$ as formatted if it has > 5 and < $|E^j|$ formatted cells.
This yields another 683K tasks, out of which we randomly sample 100K for evaluation. 
Table~\ref{tab:benchmark-tasks} shows a summary of these tasks (2).

\begin{table}[tb]
\small
\centering
\caption{
Aggregate properties of benchmark problems divided by the column data type.  \# Cells is the average number of cells in the column, and \# Formatted is the average number of cells which contain user formatting in the column.
Rule depth is defined as the tree depth of the abstract syntax tree produced by parsing the rule using our grammar.}
\label{tab:benchmark-tasks}
\begin{tabular}{lrrrr}
\toprule
  Type    & Rules  &  \# Cells & \# Formatted & Rule Depth \\ \midrule
 \multicolumn{5}{l}{(1) Tables with CF} \\ \midrule
  Text    & 13.81 K     &  107.5    & 32.1         & 2.3 \\
  Numeric & 9.32 K     &  184.8    & 111.2         & 1.8 \\
  Date    & 1.87 K     &  73.3    & 23.5         & 1.7 \\
 \textbf{Total}    & \textbf{25 K}    &  \textbf{133.7}    & \textbf{60.9}         & \textbf{2.1} \\ \midrule 
 \multicolumn{5}{l}{(2) Tables with manual formatting} \\ \midrule
  Text    & 42.4 K     &  53.6    & 17.4        & -- \\
  Numeric & 47.1 K     &  89.2    & 21.9        & -- \\
  Date    & 10.5 K     &  34.1    & 10.2          & -- \\ 
  \textbf{Total}    & \textbf{100 K}    &  \textbf{68.3}    & \textbf{49.5}         & \textbf{--} \\ \bottomrule
 \end{tabular}
\end{table}

\subsection{Metrics} \label{sec:metrics}

In this section we describe the metrics used to evaluate quality and diversity of conditions and quality of formats.

\subsubsection{Condition quality}

To evaluate a learned rule $C_L$ against a user-written rule $C_U$, we use three popular existing metrics in this domain \cite{cornet, format5}: \textbf{exact match}, \textbf{sketch match}, and \textbf{execution match}.
Exact match is a syntactic match between a learned rule and the user-written rule, with tolerance for white space and argument orderings that do not impact execution.
Sketch match is a relaxed syntactic match that only looks at the operators and rule structure between a learned rule and the user-written rule, with tolerance for differences in arguments.
Execution match consists of executing two rules and comparing the outcomes produced ($\forall E^j \in \mathbf{E}: C_L(E^j) = C_{U}(E^j)$).
In addition to capturing the fact that different rules can produce the same execution, execution match allows us to evaluate against baselines that directly predict formats, rather than produce rules.
This distinction between exact, sketch and execution match has been made in previous work in formatting~\cite{cornet, format5} and related areas, such as text to code~\cite{synchromesh, shellCodes}.

\begin{example}
    \textsf{TextEquals(\cststr{A}) $\wedge$ TextEquals(\cststr{B})} and \textsf{TextEquals(\cststr{B}) 
$\wedge$ TextEquals(\cststr{A})} are an exact match as the rules are semantically equivalent.
    \textsf{TextStartsWith(\cststr{D})} and \textsf{TextContains(\cststr{D})} are not an exact match because the rules are not equivalent.
    These may be an execution match on a column that only has \cststr{D} at the start of values.
\end{example}

\subsubsection{Condition diversity}
To evaluate diversity in conditions, we consider three pairwise metrics: \textbf{average} \textbf{edit}, \textbf{embedding} and \textbf{execution distance}.
Edit and embedding distance measure the diversity in the rule space while execution distance measures the diversity in the output format space.
Edit distance is a token level edit distance between the generated conditions, averaged over all pairwise combinations of suggestions.
Embedding distance is the average of all pairwise cosine similarities between embeddings of conditions rendered as code.
We use CodeBERT \cite{CodeBERT} to compute the embedding.
Execution distance computes diversity by considering the output of condition $C$ on a table as a boolean vector $[C_L(E^j) \mid i = 1 \rightarrow n]$.
We compute the pairwise Hamming distance over these vectors.
These metrics for diversity have previously been used to evaluate code generation systems in literature \cite{codefusion}.

\subsubsection{Format quality}
To evaluate the learner formats against the user-added formats, we consider two metrics: \textbf{color match} and \textbf{property match}.
Color match is an approximate, qualitative comparison between the format suggested and the format applied by the user.
To match colors, we map a color to the closest extended CSS color name \cite{html-web-colors} using euclidean distance between the RGB color values ($\in \mathbb{R}^3$) for both the predicted and ground truth color and match the CSS colors.
Property match is a relaxed matching which only checks if the same format properties $\subseteq F$ have been modified, without considering their values.
Property match allows us to understand the high level formatting properties and is often more indicative, since matching exact colors and fonts might be too specific and not reveal the true effectiveness of the suggestion.

\begin{example}
    \textsf{Fill(\#FF0000)} and \textsf{Fill(\#FF0011)} are an exact match as they both map to HTML color Red. On the other hand, \textsf{FontStyle(bold)} and \textsf{FontStyle(italic)} are an intent match since they have the same formatting operator (\textsf{FontStyle}) but different arguments.
\end{example}


\subsection{Baselines} \label{sec:baselines}

As we introduce the problem of predictively suggesting conditional formatting rules, there are no existing systems that perform this task.
We therefore adapt a variety of approaches related to this problem, some of which have been previously applied to conditional formatting learning.
Six approaches are symbolic, five of which are able to generate rules.
Symbolic baselines only generate conditions---no formats.
Three neural approaches cast conditional formatting as cell classification, and we consider different baseline models and cross-attention mechanisms.
Four neural baselines cast it as text generation with large language models.
Neural baselines generate both conditions and formats.
The following sections describe these baselines in more detail.
Implementation details are in Appendix~\ref{app-baselines}.

\subsection{Symbolic}

\subsubsection{Greedy Enumeration}

Greedy enumeration uses the same predicate generation as \system{} and then performs a beam search with beam width 10 and maximum depth of 5, and use the \textsc{Cornet} ranker to drive the search and select the best condition.

\subsubsection{\textsc{Cornet}}

\textsc{Cornet} \cite{cornet} is a system that learns conditional formatting rules by example.
We adapt \textsc{Cornet} to suggest CF rules without examples by replacing its semi-supervised clustering with an unsupervised clustering among two classes (formatted and non-formatted).
We did not see improvements with the iterative tree learner and therefore learn a single decision tree.

\subsubsection{Clustering}
Conditional formatting can be treated as a cell clustering problem where clusters denote the formatting groups.
We use both state-of-the-art symbolic \cite{symbolic_clustering} and neural \cite{neural_clustering} clustering techniques to split the data in groups with different formatting.
For neural clustering, the values are embedded before clustering.
Since, \textsc{Cornet} also uses clustering in its pipeline and our results show that both neural and symbolic clustering alone perform significantly worse than \textsc{Cornet}, we omit these results from the main paper.
The neural clustering results can be found in the Appendix.

\subsection{Neural}

We consider two categories of neural baselines, based on the base models.
The first category are models that learn table representations for downstream table intelligence tasks, like question-answering or cell and table type classification.
The second category are pre-trained, auto-regressive language models, which we both fine-tune and prompt to generate rules and formats over tables.

\subsubsection{TAPAS and TUTA}
TAPAS~\cite{TAPAS} is a table encoding model trained for sequential question answering (SQA) and TUTA~\cite{TUTA} is a tree-based transformer model pre-trained on multiple table tasks, such cell type classification (CTC).
We apply it to conditional formatting by using it to encode the input column and getting an embedding for each cell.
To generate conditions, use a classification head (dense model) on top of these cell embeddings that predicts formatted or not formatted.
To generate formats, use a generation head (CodeT5+ decoder) on top of the cell embeddings.


\subsubsection{\textsc{FormaT5}}

\textsc{FormaT5}~\cite{format5} is trained to generate conditional formatting rules from natural language and examples.
We remove the natural language condition but keep the abstention objective, which teaches the model to emit a placeholder token \textsf{<?>} if it cannot confidently predict a value token.
A separate model is used to predict values for these placeholders.
For example, the first model predicts \textsf{Fill(<?>)} and the second model predicts that \textsf{<?>} should be \textsf{\#FF0000}.
This yields better values predictions, because the second model can already look at the whole condition.


\subsubsection{GPT-4 \cite{openai2024gpt4}, CodeT5+ \cite{codet5+}, StarCoder \cite{starcoder} and CodeLlama \cite{codellama}}

We use open-source code generation models (CodeT5+, StarCoder and CodeLlama) through prompting and fine-tuning, and a closed-source language model (GPT-4) through prompting.
In both cases, the model is tasked with generating both the condition and the formatting.
The model prompt consists of the table information and 5 few-shot examples from the training corpus picked using CodeBERT \cite{CodeBERT} similarity.
The table is encoded as header, column types and sample values, similar to \textsc{FormaT5} \cite{format5}.
To improve performance of prompting these models, recent work has highlighted the benefit of using chain-of-thought \cite{chain-of-thought}, tree-of-thought \cite{tree-of-thought}, and program-of-thought \cite{program-of-thought}.
We find that in our experiments, program-of-thought works consistently better than the other variants and we thus only report results for program-of-thought for brevity.
The other variant results can be found in the appendix.
\section{Evaluation}
We perform experiments to answer the following questions:

\begin{itemize}[align=left]
    \item[\bfseries Q1.] Do conditions and formats generated by \system{} match those applied by users?
    \item[\bfseries Q2.] Can \system{} generate sufficient, interesting and diverse rules?
    \item[\bfseries Q3.] How do different design decisions and properties impact performance of \system{}, and are its system requirements?
\end{itemize}



\begin{table*}[htb]
\small
\centering
\caption{Comparison of \system{} with neural and symbolic baselines on learning conditions. We report sketch, exact and execution match for top-1, 3 and 5 suggestions generated.
``Rules'' denotes if an approach generates symbolic rules. \system{} outperforms neural and symbolic baselines across all execution, sketch and exact rule match.}
\label{tab:baseline}
\begin{tabularx}{\textwidth}{XXlrrrrrrrrr}
\toprule
\multicolumn{3}{c}{\textbf{System description}} & \multicolumn{3}{c}{\textbf{Execution match}} & \multicolumn{3}{c}{\textbf{Exact match}} & \multicolumn{3}{c}{\textbf{Sketch match}} \\ \cmidrule(r){1-3} \cmidrule(l){4-6} \cmidrule(l){7-9} \cmidrule(l){10-12}
\textbf{Name} & \textbf{Technique} & \textbf{Rules} & \multicolumn{1}{c}{\textbf{Top 1}} & \multicolumn{1}{c}{\textbf{Top 3}} & \multicolumn{1}{c}{\textbf{Top 5}} & \multicolumn{1}{c}{\textbf{Top 1}} & \multicolumn{1}{c}{\textbf{Top 3}} & \multicolumn{1}{c}{\textbf{Top 5}} & \multicolumn{1}{c}{\textbf{Top 1}} & \multicolumn{1}{c}{\textbf{Top 3}} & \multicolumn{1}{c}{\textbf{Top 5}} \\ \midrule
Greedy Enumeration & Symbolic & Yes & 11.8\% & 18.5\% & 25.6\% & 6.0\% & 9.7\% & 12.0\% & 36.8\% & 41.2\% & 53.3\% \\
Clustering & Symbolic & No & 13.5\% & 15.6\% & 17.4\% & -- & -- & -- & -- & -- & -- \\
\textsc{Cornet} & Symbolic & Yes & 24.3\% & 29.9\% & 39.0\% & 12.3\% & 15.7\% & 18.3\% & \underline{51.9\%} & 58.9\% & 65.8\% \\
TAPAS & Neural & Yes & 19.6\% & 23.4\% & 31.5\% & 9.9\% & 12.3\% & 14.7\% & 38.8\% & 46.1\% & 57.0\% \\
TUTA & Neural & No & 24.1\% & 30.1\% & 37.8\% & -- & -- & -- & -- & -- & -- \\
\textsc{FormaT5} & Neural & Yes & 29.4\% & 37.5\% & 57.0\% & 14.9\% & 19.6\% & 26.7\% & 47.8\% & 62.4\% & 71.9\% \\
CodeT5+ & Neural & Yes & 28.5\% & 35.5\% & 56.6\% & 14.4\% & 18.6\% & 26.5\% & 42.5\% & 57.8\% & 72.5\% \\
StarCoder & Neural & Yes & 28.3\% & 35.5\% & 55.0\% & 14.3\% & 18.6\% & 25.7\% & 42.9\% & 59.3\% & 71.7\% \\
CodeLlama & Neural & Yes & 27.1\% & 35.2\% & 53.1\% & 13.7\% & 18.4\% & 24.9\% & 41.3\% & 57.6\% & 70.4\% \\
GPT4 & Neural & Yes & \underline{31.2\%} & \underline{39.1\%} & \underline{58.3\%} & \underline{15.8\%} & \underline{20.5\%} & \underline{27.3\%} & 46.5\% & \underline{64.2\%} & \underline{75.0\%} \\
\textbf{\system{}} & \textbf{Neuro-symbolic} & \textbf{Yes} & \textbf{36.7\%} & \textbf{46.8\%} & \textbf{64.3\%} & \textbf{18.6\%} & \textbf{24.5\%} & \textbf{30.1\%} & \textbf{54.2\%} & \textbf{73.3\%} & \textbf{84.8\%} \\ \bottomrule
\end{tabularx}
\end{table*}

\subsection{Q1: Suggestion Performance}

We study the performance of learning conditions and formatting separately, and end-to-end performance of learning both.

\subsubsection{Conditions}
Table~\ref{tab:baseline} presents an overview of our results for condition learning.
\system{} outperforms symbolic and neural baselines on exact, sketch and execution match metrics.
Symbolic methods perform worse than methods with a neural component.
This neural component helps to understand the semantics of the data.
\textsc{Cornet} obtains second highest sketch match in top-1 but scores lower in execution match, which indicates that semantics are important to determine the right constant values.
\system{} combines neural and symbolic components to extract predicate with interesting constants and then ranks them to obtain best of both worlds.

\begin{figure}[tb]
    \centering
    \includegraphics[width=\columnwidth]{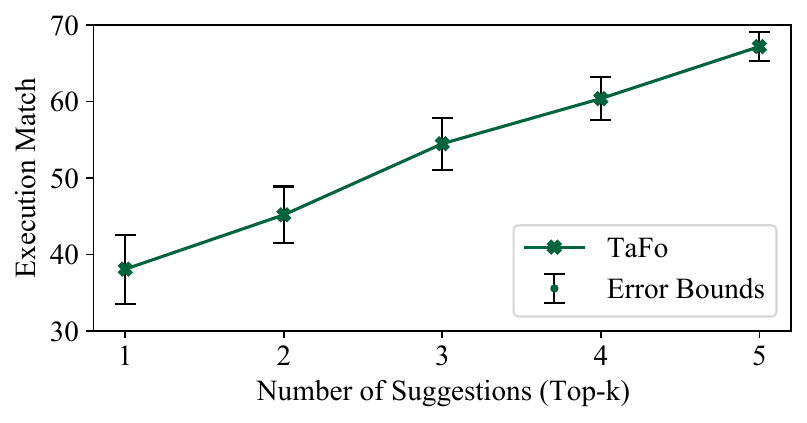}
    \caption{\system{} performance on tasks with manual formatting. We show execution match accuracy for increasing number of samples generated (top-$k$).}
    \label{fig:eval-manual-formatting}
\end{figure}

In addition to evaluating \system{} on conditions written by users, we also investigate the extent to which it can learn conditions for formatting that users carried out manually.
Since there is no rule, we report execution match.
Figure~\ref{fig:eval-manual-formatting} shows the execution match for \system{} with increasing number of suggestions generated (top-$k$).
We find that \system{} could have automated 55\% of column formatting for the users with just three suggestions.



\subsubsection{Formatting}

\begin{table}[tb]
\small
\centering
\caption{(Formatting Identifier) Comparison of \system{} with neural and symbolic baselines. We report exact and property match for top-1, 3 and 5 suggestions generated.
\system{} outperforms all baselines accross both exact and property match.
}
\label{tab:format-perf}
\begin{tabular}{lrrrrrr}
\toprule
\multicolumn{1}{c}{\textbf{System}} & \multicolumn{3}{c}{\textbf{Exact match}} & \multicolumn{3}{c}{\textbf{Property match}} \\ \cmidrule(r){1-1} \cmidrule(l){2-4} \cmidrule(l){5-7}
\textbf{Name} & \multicolumn{1}{c}{\textbf{Top 1}} & \multicolumn{1}{c}{\textbf{Top 3}} & \multicolumn{1}{c}{\textbf{Top 5}} & \multicolumn{1}{c}{\textbf{Top 1}} & \multicolumn{1}{c}{\textbf{Top 3}} & \multicolumn{1}{c}{\textbf{Top 5}} \\ \midrule
TAPAS & 4.3\% & 7.1\% & 9.0\% & 8.6\% & 17.7\% & 23.4\% \\
FormaT5 & 9.5\% & 14.3\% & 16.2\% & 19.1\% & 35.7\% & 42.1\% \\
CodeT5+ & 8.1\% & 10.1\% & 11.9\% & 16.3\% & 25.2\% & 31.0\% \\
StarCoder & 7.7\% & 9.9\% & 11.5\% & 15.5\% & 24.7\% & 29.9\% \\
CodeLlama & 7.1\% & 9.3\% & 10.8\% & 14.3\% & 23.2\% & 28.1\% \\
GPT4 & 8.4\% & 10.5\% & 12.2\% & 16.9\% & 26.2\% & 31.7\% \\
\textbf{\system{}} & \textbf{13.2\%} & \textbf{18.9\%} & \textbf{21.3\%} & \textbf{26.5\%} & \textbf{47.2\%} & \textbf{55.4\%} \\ \bottomrule
\end{tabular}
\end{table}

As formatting is a visual task, we evaluate \system{} and baselines on their ability to generate the formats that the users added in their tables.
For this experiment, we only match the format and ignore the condition.

Table~\ref{tab:format-perf} shows the exact and property match for top-1, 3 and 5 suggestions for \system{} and baselines. 
We find that \system{} performs consistently better than baselines on both exact match (+4.8--9.1\%) and property match (+9.4--23.7\%) over the best baseline.

Qualitative analysis of the baseline performance reveals insights about the common mistakes made by these systems.
Neural models like GPT-4 tend to prefer extreme colors like ``red'' and ``green'' for boolean operations.
On examination of ground truths we find that in a lot of these cases, however, the actual user preference is a mild color like ``pink'' or ``yellow''.
\system{} is able to correctly generate the right format, as these patterns are prevalent in the corpus.

\subsubsection{End-to-end}

\begin{figure}
    \centering
    \includegraphics[width=\columnwidth]{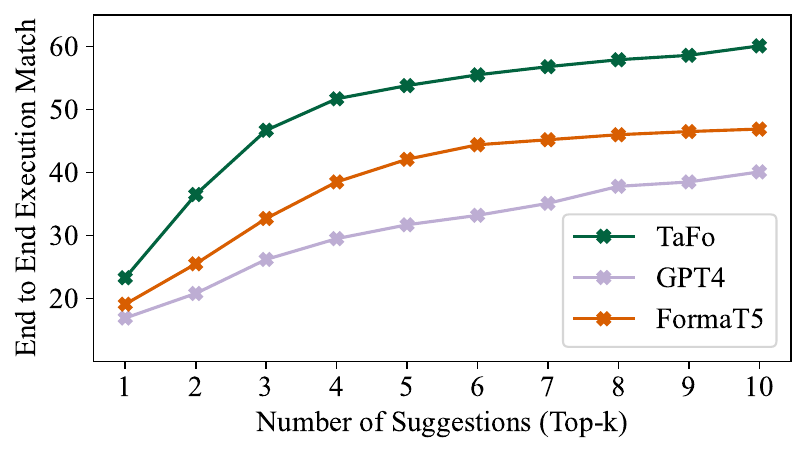}
    \caption{End to end execution match for \system{}, the best baseline (GPT-4) and the current state-of-the-art in table formatting (\textsc{FormaT5}) against increasing number of suggestions (Top-$k$). \system{} outperforms both baselines}
    \label{fig:eval-end-to-end}
\end{figure}

We also evaluate end-to-end performance of \system{}, which considers both the rule trigger and format together.
We consider end-to-end execution match as (condition execution match) $\wedge$ (format color match).
Figure~\ref{fig:eval-end-to-end} shows the end-to-end execution match for \system{} and baselines on increasing number of suggestions.
We find that \system{} performs consistently better than the next best baselines (GPT-4 and FormaT5).
\system{} gets an end-to-end execution match of 50\% with just three suggestions.

\subsection{Q2: Condition Analysis}\label{ssec:exp-analysis}

It is important that we generate various (completeness and diversity) and interesting (complexity and coverage) conditions.
We study these properties in the following paragraphs.

\paragraph{Completeness}

\begin{figure}[tb]
    \centering
    \begin{subfigure}{.49\columnwidth}
        \centering
        \includegraphics[width=\columnwidth]{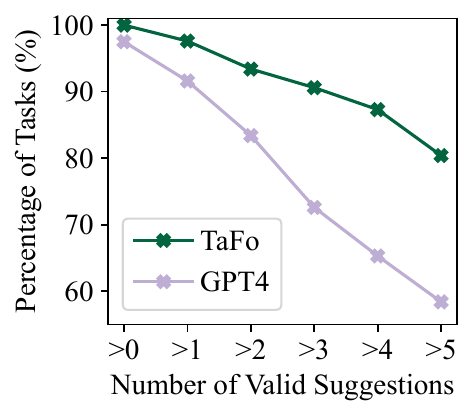}
        \caption{Completeness}
        \label{fig:completion}
    \end{subfigure}
    \begin{subfigure}{.49\columnwidth}
        \includegraphics[width=\columnwidth]{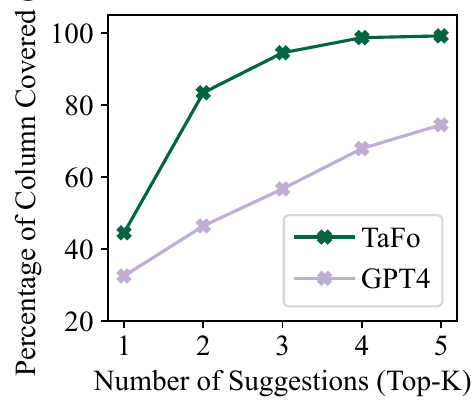}
        \caption{Coverage}
        \label{fig:coverage}
    \end{subfigure}
    \caption{(a) Completeness: Percentage tasks with greater than $k$ suggestions; and (b) Coverage: Percentage of the column covered by at least one rule in the suggestions for \system{} and GPT-4. We find that \system{} generates three or more suggestions for over 90\% of the benchmark tasks and achieves a column coverage of over 80\% with just two suggestions.}
    \label{fig:completion_coverage}
\end{figure}

Figure~\ref{fig:completion} shows the number of tasks for which at least $> k$ suggestions are made for increasing $k$, for both \system{} and GPT-4.
This corresponds to an upper bound of the recall after $k$ suggestions.
Because we combine different methods of generation, among which exhaustively generating predicates, \system{} obtains a higher upper bound on recall.


\paragraph{Coverage}

Figure~\ref{fig:coverage} shows percentage of cells in a column that are covered by any condition for increasing number of suggestions, for \system{} and GPT-4.
Because \system{} ranks suggestions based on their execution, it covers significantly more cells.
This is especially noticeable after the second suggestion, which upon execution, affects completely different cells than the first suggestion.
With just 2 suggestions, \system{} achieves a column coverage of over 80\%.

\paragraph{Complexity}

\begin{figure}[tb]
    \centering
    \begin{subfigure}{.49\columnwidth}
        \centering
        \includegraphics[width=\columnwidth]{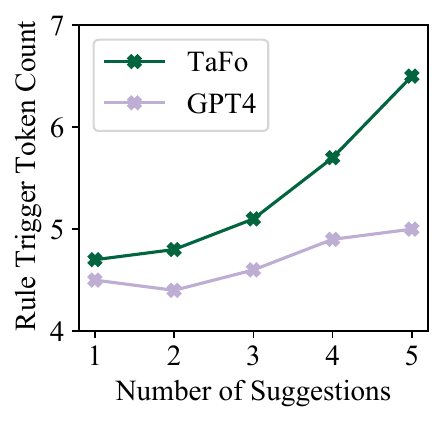}
        \caption{Token Count}
        \label{fig:complexity-tokens}
    \end{subfigure}
    \begin{subfigure}{.495\columnwidth}
        \includegraphics[width=\columnwidth]{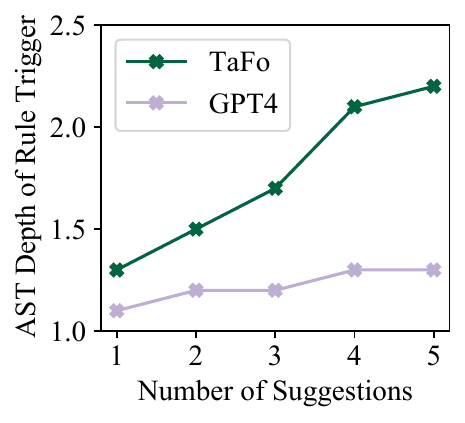}
        \caption{AST Depth}
        \label{fig:complexity-ast}
    \end{subfigure}
    \caption{Rule trigger $t$ complexity for suggestions generated by \system{} and the best baseline (GPT4). We show (a) Token Count: average number of tokens in the suggestions; and (b) AST Depth: average depth of rule trigger in the AST. We find that the complexity of conditions generated by \system{} increases with the number of suggestions as most of the low complexity rules have similar execution.}
    \label{fig:complexity}
\end{figure}

Figure~\ref{fig:complexity} shows how the complexity evolves for generating more conditions based on two metrics: (a) the number of tokens in the condition, with tokens defined as predicates and arguments; and (b) depth of the syntax tree of the condition.
As opposed to GPT-4, \system{} yields more complex conditions as the number of conditions increases.
This happens for two reasons: recombining simpler conditions (with predicates at the lowest level) into more complex conditions and ranking these based on execution to ensure that complex conditions are also chosen.


\paragraph{Diversity}

Table~\ref{tab:eval-diversity} summarizes condition diversity metrics of \system{} and baselines over the top 5 suggestions.
We only consider the best performing baseline from each category---the remainder of results is in the Appendix.
Because of ranking with execution, \system{} achieves highest diversity across all metrics.
Interestingly, neural models (GPT-4, CodeT5+) struggle with diversity, despite good overall performance (Table~\ref{tab:baseline}).
Low-diversity generations is a known issue in large language models \cite{10.1145/3597307, codefusion}.
\textsc{FormaT5} and \textsc{Cornet} also ranks candidates and achieves high rule diversity, but they do not consider differences in execution across rules.

\begin{table}[tb]
\small
\centering
\caption{Diversity metrics for top-$5$ suggestion generation for \system{} and baselines. We report average pairwise edit distance, embedding similarity and Hamming distance of the output formatting on execution. Arrows in the header indicate if higher ($\uparrow$) or lower ($\downarrow$) value is better.}
\label{tab:eval-diversity}
\begin{tabularx}{.99\columnwidth}{Xrrr}
\toprule
 & \multicolumn{3}{c}{Diversity Metric} \\ \cmidrule{2-4}
\textbf{System} & \textbf{Edit Distance ($\uparrow$)} & \textbf{Embedding ($\downarrow$)} & \textbf{Execution ($\uparrow$)} \\ \midrule
TAPAS   & 6.2  & 0.95 & 0.27 \\
\textsc{Cornet}  & 9.2  & 0.94 & 0.33 \\
\textsc{FormaT5} & 10.4 & 0.91 & 0.37 \\
CodeT5+ & 6.5  & 0.96 & 0.25 \\
GPT-4    & 7.6  & 0.94 & 0.31 \\
\textbf{\system{}} & \textbf{14.3} & \textbf{0.87} & \textbf{0.65} \\
\bottomrule
\end{tabularx}
\end{table}

\subsection{Q3: Design Choices and Input Properties}\label{ssec:design}

We evaluate the impact of different design decisions and input properties on the performance of \system{}.

\paragraph{Symbolic $\leftrightarrow$ neural $\leftrightarrow$ neuro-symbolic}

\begin{figure}
    \centering
    \includegraphics[width=\columnwidth]{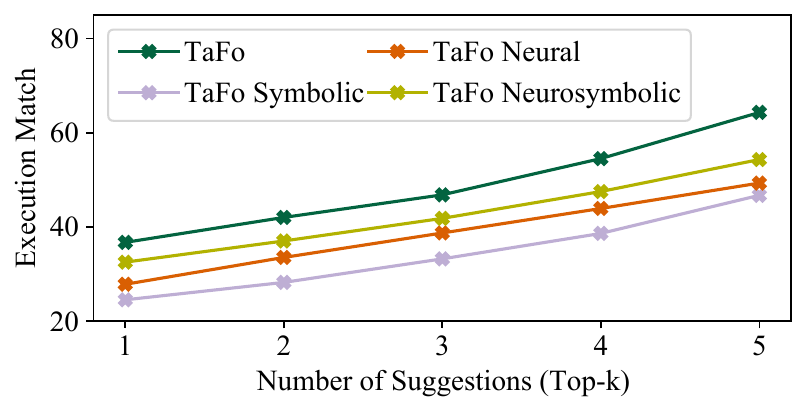}
    \caption{
    Execution match for condition learning for \system{} and its neural, symbolic and neurosymbolic only variants for increasing number of suggestions (Top-$k$). \system{} outperforms all ablations, especially with higher number of suggestions, which highlights the value of combining the neural, symbolic and neurosymbolic techniques. Further, the neurosymbolic variant outperforms both neural and symbolic variants.}
    \label{fig:eval-neural-v-symbolic}
\end{figure}

The neural and symbolic components of \system{} can be decoupled and work independently---see Section~\ref{sec:method-trigger}.
Figure~\ref{fig:eval-neural-v-symbolic} shows the execution match accuracy of condition learning for \system{} and its purely neural, symbolic and neurosymbolic ablations.
We find that \system{} outperforms all ablations by 4.5\%--16.7\% and 5.8\%--21.8\% respectively showing that different condition learners specialize on different tasks thus indicating the value of having multiple condition learners. Further, the neurosymbolic ablation outperforms both the purely neural and symbolic, highlighting the value of sharing reasoning paths.

\paragraph{Condition ablations}

\begin{table}[tb]
\small
\centering
\caption{Execution match for condition learning for different ablations of \system{}. Each ablated component is denoted by `--'. Symbolic properties (row 2) and ranking (row 1) have the highest impact on performance. Neuro-symbolic interaction also affects performance of \system{}, improving over directly combining candidates from symbolic and neural variants.
}
\label{tab:eval-ablation}
\begin{tabularx}{.99\columnwidth}{Xrrr}
\toprule
\textbf{System} & \textbf{Top-1} & \textbf{Top-3} & \textbf{Top-5} \\ \midrule
\system{} -- execution-guided ranking & 19.8\% & 34.5\% & 56.4\% \\
\system{} -- symbolic property extraction & 17.2\% & 20.7\% & 53.4\% \\
\system{} -- LLM multi-step reasoning & 23.5\% & 38.2\% & 58.8\% \\
\system{} -- neuro-symbolic interaction & 25.8\% & 40.0\% & 59.6\% \\
\textbf{\system{}} & \textbf{36.7\%} & \textbf{46.8\%} & \textbf{64.3\%} \\
\bottomrule
\end{tabularx}
\end{table}

We ablate different components to understand their impact on the performance of \system{}.
We ablate execution guided ranking by considering the original order of rules returned by the trigger learner.
We ablate symbolic property extraction by directly running enumeration and LLM generation without these properties.
We ablate LLM generation by not using multi-step reasoning.
Table~\ref{tab:eval-ablation} shows execution match results for condition learning for top-1, 3 and 5 suggestions for these ablations.
We find that symbolic property extraction has the biggest impact on performance (-19.5\% in top-1). This is expected since the properties provide building blocks for the symbolic and neural models to reason about the data.
Execution-guided ranking significantly affects the top-5 performance (-20.4\%) since not selecting diverse samples reduces coverage and causes repetition in the suggestions.

\paragraph{Format ablations}

\begin{table}[tb]
\small
\centering
\caption{Intent match for different ablations of \system{} format learner ($F_{\system{}}$). Each ablated component is denoted by `--'. Corpus retrieval has the highest impact on the overall performance of \system{} format learner showing that similar data have similar format preferences.}
\label{tab:eval-format-ablation}
\begin{tabularx}{.99\columnwidth}{Xrrr}
\toprule
\textbf{System} & \textbf{Top-1} & \textbf{Top-3} & \textbf{Top-5} \\ \midrule
$F_{\system{}}$ -- Corpus Retrieval & 17.5\% & 28.1\% & 34.8\% \\
$F_{\system{}}$ -- Sheet Properties & 23.1\% & 40.0\% & 48.2\% \\
$F_{\system{}}$ -- Format Grounding & 21.4\% & 36.6\% & 44.4\% \\
\textbf{$F_{\system{}}$} & \textbf{26.5\%} & \textbf{47.2\%} & \textbf{55.4\%} \\
\bottomrule
\end{tabularx}
\end{table}

Table~\ref{tab:eval-format-ablation} shows the performance of learning formats with different components removed.
We find that the corpus retrieval has the most significant impact on performance (-19.1\% -- -9.0\%). This is consistent with recent work \cite{chen2024autoformula} which finds that mining formula components and other artifacts from similar tables boosts performance since similar data has similarity in associated artifacts like formulas and formats.
Not grounding the format impacts performance (-4.5\%).
Furthermore, not using sheet properties to align formats to the current sheet also affects performance (-3.4\%).

\paragraph{Sample size of rows}

\begin{figure}[tb]
\centering
\includegraphics[width=\columnwidth]{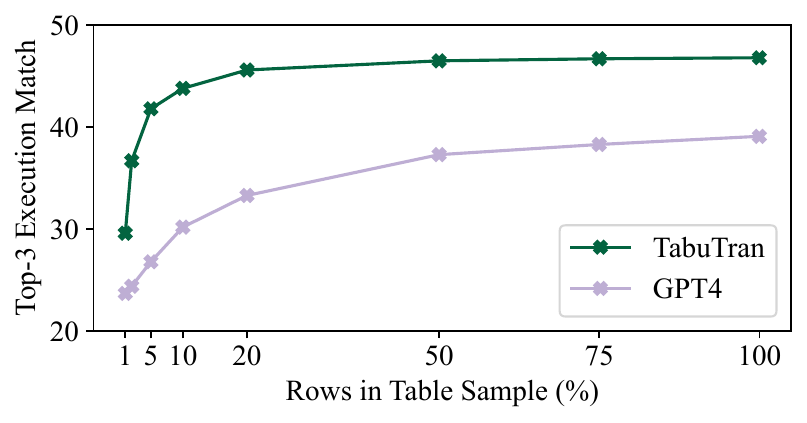}
\caption{
Top-3 suggestion execution match for condition learning for \system{} and best baseline (GPT-4) when using between 1\% and 100\% of rows. We sample rows randomly from tables with > 1000 rows. \system{} achieves over 40\% execution match with just 10\% of rows while GPT-4 only achieves 30\%.}
\label{fig:eval-samplesize}
\end{figure}

The quality of suggestions is directly impacted by the amount of data available, however, due to practical constraints, the entire table data might not be readily available.
For example, spreadsheet clients often store a subset of data on the user viewport (active screen) which is updated as the viewport moves -- for large spreadsheets this can be a small fraction of their total data. To be practically useful, \system{} must be able to work with table subsets without a significant drop in suggestion quality.
Figure~\ref{fig:eval-samplesize} shows the execution match accuracy when including from 1\% to 100\% of rows in the table sample.
We find that the performance drops as the sample size is reduced.
However, with just 5\% rows, \system{} gets an execution match of 42.8\% with 3 suggestions per task.

\paragraph{Data type of rule}

\begin{figure}[tb]
\centering
\includegraphics[width=\columnwidth]{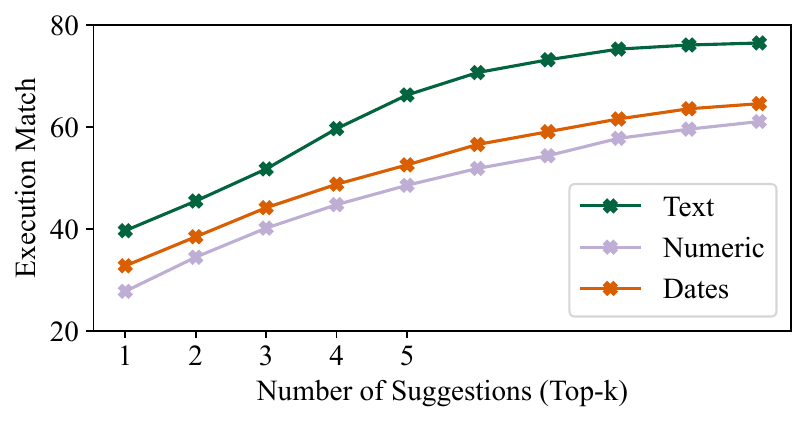}
\caption{
Execution match for condition learning for \system{} over increasing number of suggestions (Top-$k$) for different column data types. \system{} has higher
accuracy for Text columns. Numeric and Date columns
keep increasing with more suggestions, due to the larger search space of rules.}
\label{fig:datatype}
\end{figure}

We analyze the condition learning performance of \system{} based on the data types of the suggestions. Figure~\ref{fig:data-type} shows the execution match for different rule data types with increasing number of generation size (Top-$k$). 
We find that \system{} has the highest performance on text columns and numeric and date columns have lower execution match with fewer suggestions and it improves at a lower rate compared to text columns on increasing the number of suggestions. This is inline with what previous work in spreadsheet formatting~\cite{cornet} have observed due to the numeric and date rules having an infinitely large search space for arguments.

\paragraph{Execution Time and Memory}

\begin{table}[tb]
\small
\centering
\caption{Total disk space and average memory used in megabytes for inference over benchmark tasks for \system{} and baselines.
GPU usage for \system{} is not included as it uses GPT4 via API.
\system{} uses GPT4 via API and the latency reported includes network latency.
Symbolic variant of \system{} is lightweight with latency of only 189.3 ms.
}
\label{tab:eval-space-time}
\begin{tabularx}{.99\columnwidth}{Xrrrr}
\toprule
\textbf{System} & \textbf{Latency} & \textbf{Disk} & \textbf{CPU} & \textbf{GPU} \\ \midrule
Greedy Enumeration & 781.3 & 3.1 & 103.4 & 0 \\
\textsc{Cornet} & 254.3 & 2.9 & 61.4 & 0 \\
TUTA & 2194.4 & 741.5 & 7.6 & 1432.8 \\
TAPAS & 1832.5 & 412.4 & 9.1 & 1713.2 \\
\system{} & 1218.5 & 4.5 & 8.3 & -- \\
\system{} Symbolic & 189.3 & 4.5 & 8.9 & 0 \\
\bottomrule
\end{tabularx}
\end{table}

We also evaluate the time and memory resources required by \system{} and baselines.
Table~\ref{tab:eval-space-time} shows the average latency, total amount of disk space required to store a system, and the average CPU and GPU memory used for prediction over benchmarks for
\system{} and baselines.
We also report the latency and memory of \system{} for just the symbolic ablation which does not require LLM resources.
Section~\ref{sec:method-symbolic} describes the symbolic variant of \system{} designed for low-resource environments.
Since \system{} makes LLM calls over an API (GPT-4), we do not report GPU memory for \system{} and the latency reported includes network latency. Appendix contains the tokens and cost analysis for \system{} and discussion over replacing the LLM with a smaller and cheaper model.


\section{Conclusion}
In this paper, we introduced the novel problem of 
learning conditional formatting rules without user specification.
We proposed \system{}, a system that learns such data-dependent formatting suggestions with rule and format based on tabular context.
We evaluate \system{}, on a benchmark of 105K CF tasks extracted from 1.8 million real Excel spreadsheets and evaluate quality, completeness, diversity and complexity of the suggestions.
We compare \system{} to both symbolic and neural approaches on this benchmark, for the table formatting task and also adapt techniques from other tasks in this domain,
and find that \system{} performs significantly better than baselines.
We evaluate \system{} on varying input properties to simulate practical challenges like availability of data and latency constraints.
We experiment with the components of \system{} and design choices made, analyzing the impact they have on overall performance.
Finally, we evaluate \system{} on manually formatted tables which contain formatting but no rules and find that \system{} automates over 60\% of these tasks.
\system{} opens future work for other tabular predictive tasks like cleaning, transformation, visualization and querying. Furthermore, \system{} highlights the importance of corpus grounding and neurosymbolic interaction in recommender systems for tables.

\bibliographystyle{ACM-Reference-Format}
\bibliography{references}


\begin{thebibliography}{53}


\ifx \showCODEN    \undefined \def \showCODEN     #1{\unskip}     \fi
\ifx \showISBNx    \undefined \def \showISBNx     #1{\unskip}     \fi
\ifx \showISBNxiii \undefined \def \showISBNxiii  #1{\unskip}     \fi
\ifx \showISSN     \undefined \def \showISSN      #1{\unskip}     \fi
\ifx \showLCCN     \undefined \def \showLCCN      #1{\unskip}     \fi
\ifx \shownote     \undefined \def \shownote      #1{#1}          \fi
\ifx \showarticletitle \undefined \def \showarticletitle #1{#1}   \fi
\ifx \showURL      \undefined \def \showURL       {\relax}        \fi
\providecommand\bibfield[2]{#2}
\providecommand\bibinfo[2]{#2}
\providecommand\natexlab[1]{#1}
\providecommand\showeprint[2][]{arXiv:#2}

\bibitem[Abramovich et~al\mbox{.}(2004)]%
        {Abramovich04spreadsheetconditional}
\bibfield{author}{\bibinfo{person}{Sergei Abramovich}, \bibinfo{person}{Stephen Sugden}, \bibinfo{person}{Sergei Abramovich}, {and} \bibinfo{person}{Stephen~J Sugden}.} \bibinfo{year}{2004}\natexlab{}.
\newblock \showarticletitle{Spreadsheet Conditional Formatting: An Untapped Resource for Mathematics Education}.
\newblock \bibinfo{journal}{\emph{Spreadsheets in Education}}  \bibinfo{volume}{1} (\bibinfo{year}{2004}), \bibinfo{pages}{85105}.
\newblock


\bibitem[Chen et~al\mbox{.}(2024)]%
        {chen2024autoformula}
\bibfield{author}{\bibinfo{person}{Sibei Chen}, \bibinfo{person}{Yeye He}, \bibinfo{person}{Weiwei Cui}, \bibinfo{person}{Ju Fan}, \bibinfo{person}{Song Ge}, \bibinfo{person}{Haidong Zhang}, \bibinfo{person}{Dongmei Zhang}, {and} \bibinfo{person}{Surajit Chaudhuri}.} \bibinfo{year}{2024}\natexlab{}.
\newblock \bibinfo{title}{Auto-Formula: Recommend Formulas in Spreadsheets using Contrastive Learning for Table Representations}.
\newblock
\showeprint[arxiv]{2404.12608}~[cs.DB]


\bibitem[Chen et~al\mbox{.}(2023)]%
        {program-of-thought}
\bibfield{author}{\bibinfo{person}{Wenhu Chen}, \bibinfo{person}{Xueguang Ma}, \bibinfo{person}{Xinyi Wang}, {and} \bibinfo{person}{William~W. Cohen}.} \bibinfo{year}{2023}\natexlab{}.
\newblock \showarticletitle{Program of Thoughts Prompting: Disentangling Computation from Reasoning for Numerical Reasoning Tasks}.
\newblock \bibinfo{journal}{\emph{Transactions on Machine Learning Research}} (\bibinfo{year}{2023}).
\newblock
\showISSN{2835-8856}
\urldef\tempurl%
\url{https://openreview.net/forum?id=YfZ4ZPt8zd}
\showURL{%
\tempurl}


\bibitem[Chen et~al\mbox{.}(2021)]%
        {spreadsheetCoder}
\bibfield{author}{\bibinfo{person}{Xinyun Chen}, \bibinfo{person}{Petros Maniatis}, \bibinfo{person}{Rishabh Singh}, \bibinfo{person}{Charles Sutton}, \bibinfo{person}{Hanjun Dai}, \bibinfo{person}{Max Lin}, {and} \bibinfo{person}{Denny Zhou}.} \bibinfo{year}{2021}\natexlab{}.
\newblock \showarticletitle{SpreadsheetCoder: Formula Prediction from Semi-structured Context}. In \bibinfo{booktitle}{\emph{Proceedings of the 38th International Conference on Machine Learning}} \emph{(\bibinfo{series}{Proceedings of Machine Learning Research}, Vol.~\bibinfo{volume}{139})}, \bibfield{editor}{\bibinfo{person}{Marina Meila} {and} \bibinfo{person}{Tong Zhang}} (Eds.). \bibinfo{publisher}{PMLR}, \bibinfo{address}{virtual}, \bibinfo{pages}{1661--1672}.
\newblock
\urldef\tempurl%
\url{https://proceedings.mlr.press/v139/chen21m.html}
\showURL{%
\tempurl}


\bibitem[Dong et~al\mbox{.}(2020)]%
        {numerical_formatting}
\bibfield{author}{\bibinfo{person}{Haoyu Dong}, \bibinfo{person}{Jinyu Wang}, \bibinfo{person}{Zhouyu Fu}, \bibinfo{person}{Shi Han}, {and} \bibinfo{person}{Dongmei Zhang}.} \bibinfo{year}{2020}\natexlab{}.
\newblock \showarticletitle{Neural Formatting for Spreadsheet Tables}. In \bibinfo{booktitle}{\emph{Proceedings of the 29th ACM International Conference on Information \& Knowledge Management}} (Virtual Event, Ireland) \emph{(\bibinfo{series}{CIKM '20})}. \bibinfo{publisher}{Association for Computing Machinery}, \bibinfo{address}{New York, NY, USA}, \bibinfo{pages}{305–314}.
\newblock
\showISBNx{9781450368599}
\href{https://doi.org/10.1145/3340531.3411943}{doi:\nolinkurl{10.1145/3340531.3411943}}


\bibitem[Ellis and Gulwani(2017)]%
        {ranking_program_feats}
\bibfield{author}{\bibinfo{person}{Kevin Ellis} {and} \bibinfo{person}{Sumit Gulwani}.} \bibinfo{year}{2017}\natexlab{}.
\newblock \showarticletitle{Learning to Learn Programs from Examples: Going Beyond Program Structure}. In \bibinfo{booktitle}{\emph{IJCAI 2017} (\bibinfo{edition}{ijcai 2017} ed.)}. \bibinfo{publisher}{IJCAI 2017}, \bibinfo{address}{Melbourne, Australia}, \bibinfo{pages}{1638--1645}.
\newblock
\urldef\tempurl%
\url{www.microsoft.com/research/publication/learning-learn-programs-examples-going-beyond-program-structure/}
\showURL{%
\tempurl}


\bibitem[Fariha and Meliou(2019)]%
        {vldb-pbe-1}
\bibfield{author}{\bibinfo{person}{Anna Fariha} {and} \bibinfo{person}{Alexandra Meliou}.} \bibinfo{year}{2019}\natexlab{}.
\newblock \showarticletitle{Example-Driven Query Intent Discovery: Abductive Reasoning Using Semantic Similarity}.
\newblock \bibinfo{journal}{\emph{Proc. VLDB Endow.}} \bibinfo{volume}{12}, \bibinfo{number}{11} (\bibinfo{date}{jul} \bibinfo{year}{2019}), \bibinfo{pages}{1262–1275}.
\newblock
\showISSN{2150-8097}
\href{https://doi.org/10.14778/3342263.3342266}{doi:\nolinkurl{10.14778/3342263.3342266}}


\bibitem[Fariha et~al\mbox{.}(2021)]%
        {vldb-pbe-2}
\bibfield{author}{\bibinfo{person}{Anna Fariha}, \bibinfo{person}{Ashish Tiwari}, \bibinfo{person}{Alexandra Meliou}, \bibinfo{person}{Arjun Radhakrishna}, {and} \bibinfo{person}{Sumit Gulwani}.} \bibinfo{year}{2021}\natexlab{}.
\newblock \showarticletitle{CoCo: Interactive Exploration of Conformance Constraints for Data Understanding and Data Cleaning}. In \bibinfo{booktitle}{\emph{Proceedings of the 2021 International Conference on Management of Data}} (Virtual Event, China) \emph{(\bibinfo{series}{SIGMOD '21})}. \bibinfo{publisher}{Association for Computing Machinery}, \bibinfo{address}{New York, NY, USA}, \bibinfo{pages}{2706–2710}.
\newblock
\showISBNx{9781450383431}
\href{https://doi.org/10.1145/3448016.3452750}{doi:\nolinkurl{10.1145/3448016.3452750}}


\bibitem[Feng et~al\mbox{.}(2020)]%
        {CodeBERT}
\bibfield{author}{\bibinfo{person}{Zhangyin Feng}, \bibinfo{person}{Daya Guo}, \bibinfo{person}{Duyu Tang}, \bibinfo{person}{Nan Duan}, \bibinfo{person}{Xiaocheng Feng}, \bibinfo{person}{Ming Gong}, \bibinfo{person}{Linjun Shou}, \bibinfo{person}{Bing Qin}, \bibinfo{person}{Ting Liu}, \bibinfo{person}{Daxin Jiang}, {and} \bibinfo{person}{Ming Zhou}.} \bibinfo{year}{2020}\natexlab{}.
\newblock \showarticletitle{{C}ode{BERT}: A Pre-Trained Model for Programming and Natural Languages}. In \bibinfo{booktitle}{\emph{EMNLP 2020}}. \bibinfo{publisher}{Association for Computational Linguistics}, \bibinfo{address}{Online}, \bibinfo{pages}{1536--1547}.
\newblock
\href{https://doi.org/10.18653/v1/2020.findings-emnlp.139}{doi:\nolinkurl{10.18653/v1/2020.findings-emnlp.139}}


\bibitem[Gulwani(2011)]%
        {flashfill}
\bibfield{author}{\bibinfo{person}{Sumit Gulwani}.} \bibinfo{year}{2011}\natexlab{}.
\newblock \showarticletitle{Automating String Processing in Spreadsheets using Input-Output Examples}. In \bibinfo{booktitle}{\emph{PoPL'11, January 26-28, 2011, Austin, Texas, USA}}. \bibinfo{publisher}{Association for Computing Machinery}, \bibinfo{address}{New York, NY, USA}, \bibinfo{pages}{317–330}.
\newblock
\urldef\tempurl%
\url{https://www.microsoft.com/en-us/research/publication/automating-string-processing-spreadsheets-using-input-output-examples/}
\showURL{%
\tempurl}


\bibitem[Herzig et~al\mbox{.}(2020)]%
        {TAPAS}
\bibfield{author}{\bibinfo{person}{Jonathan Herzig}, \bibinfo{person}{Paweł~Krzysztof Nowak}, \bibinfo{person}{Thomas Müller}, \bibinfo{person}{Francesco Piccinno}, {and} \bibinfo{person}{Julian~Martin Eisenschlos}.} \bibinfo{year}{2020}\natexlab{}.
\newblock \showarticletitle{Tapas: Weakly Supervised Table Parsing via Pre-training}. In \bibinfo{booktitle}{\emph{Proceedings of the 58th Annual Meeting of the Association for Computational Linguistics (Volume 1: Long Papers)}}. \bibinfo{publisher}{Association for Computational Linguistics}, \bibinfo{address}{Seattle, Washington, United States}, \bibinfo{pages}{4320--4333}.
\newblock
\urldef\tempurl%
\url{https://www.aclweb.org/anthology/2020.acl-main.398/}
\showURL{%
\tempurl}


\bibitem[Hong et~al\mbox{.}(2024)]%
        {data-interpret}
\bibfield{author}{\bibinfo{person}{Sirui Hong}, \bibinfo{person}{Yizhang Lin}, \bibinfo{person}{Bang Liu}, \bibinfo{person}{Bangbang Liu}, \bibinfo{person}{Binhao Wu}, \bibinfo{person}{Danyang Li}, \bibinfo{person}{Jiaqi Chen}, \bibinfo{person}{Jiayi Zhang}, \bibinfo{person}{Jinlin Wang}, \bibinfo{person}{Li Zhang}, \bibinfo{person}{Lingyao Zhang}, \bibinfo{person}{Min Yang}, \bibinfo{person}{Mingchen Zhuge}, \bibinfo{person}{Taicheng Guo}, \bibinfo{person}{Tuo Zhou}, \bibinfo{person}{Wei Tao}, \bibinfo{person}{Wenyi Wang}, \bibinfo{person}{Xiangru Tang}, \bibinfo{person}{Xiangtao Lu}, \bibinfo{person}{Xiawu Zheng}, \bibinfo{person}{Xinbing Liang}, \bibinfo{person}{Yaying Fei}, \bibinfo{person}{Yuheng Cheng}, \bibinfo{person}{Zongze Xu}, {and} \bibinfo{person}{Chenglin Wu}.} \bibinfo{year}{2024}\natexlab{}.
\newblock \bibinfo{title}{Data Interpreter: An LLM Agent For Data Science}.
\newblock
\showeprint[arxiv]{2402.18679}~[cs.AI]


\bibitem[Hou et~al\mbox{.}(2024)]%
        {llm-ranker}
\bibfield{author}{\bibinfo{person}{Yupeng Hou}, \bibinfo{person}{Junjie Zhang}, \bibinfo{person}{Zihan Lin}, \bibinfo{person}{Hongyu Lu}, \bibinfo{person}{Ruobing Xie}, \bibinfo{person}{Julian McAuley}, {and} \bibinfo{person}{Wayne~Xin Zhao}.} \bibinfo{year}{2024}\natexlab{}.
\newblock \bibinfo{title}{Large Language Models are Zero-Shot Rankers for Recommender Systems}.
\newblock
\showeprint[arxiv]{2305.08845}~[cs.IR]


\bibitem[Hurst et~al\mbox{.}(2005)]%
        {rel_formatting_2}
\bibfield{author}{\bibinfo{person}{Nathan Hurst}, \bibinfo{person}{Kim Marriott}, {and} \bibinfo{person}{Peter Moulder}.} \bibinfo{year}{2005}\natexlab{}.
\newblock \showarticletitle{Toward tighter tables}. In \bibinfo{booktitle}{\emph{Proceedings of the 2005 ACM symposium on Document engineering}}. \bibinfo{publisher}{Association for Computing Machinery}, \bibinfo{address}{New York, NY, USA}, \bibinfo{pages}{74--83}.
\newblock


\bibitem[Joshi et~al\mbox{.}(2023)]%
        {flame}
\bibfield{author}{\bibinfo{person}{Harshit Joshi}, \bibinfo{person}{Abishai Ebenezer}, \bibinfo{person}{José Cambronero}, \bibinfo{person}{Sumit Gulwani}, \bibinfo{person}{Aditya Kanade}, \bibinfo{person}{Vu Le}, \bibinfo{person}{Ivan Radiček}, {and} \bibinfo{person}{Gust Verbruggen}.} \bibinfo{year}{2023}\natexlab{}.
\newblock \bibinfo{title}{FLAME: A small language model for spreadsheet formulas}.
\newblock
\showeprint[arxiv]{2301.13779}~[cs.PL]


\bibitem[Khatry et~al\mbox{.}(2023)]%
        {khatry2023wordscodeharnessingdata}
\bibfield{author}{\bibinfo{person}{Anirudh Khatry}, \bibinfo{person}{Joyce Cahoon}, \bibinfo{person}{Jordan Henkel}, \bibinfo{person}{Shaleen Deep}, \bibinfo{person}{Venkatesh Emani}, \bibinfo{person}{Avrilia Floratou}, \bibinfo{person}{Sumit Gulwani}, \bibinfo{person}{Vu Le}, \bibinfo{person}{Mohammad Raza}, \bibinfo{person}{Sherry Shi}, \bibinfo{person}{Mukul Singh}, {and} \bibinfo{person}{Ashish Tiwari}.} \bibinfo{year}{2023}\natexlab{}.
\newblock \bibinfo{title}{From Words to Code: Harnessing Data for Program Synthesis from Natural Language}.
\newblock
\showeprint[arxiv]{2305.01598}~[cs.DB]
\urldef\tempurl%
\url{https://arxiv.org/abs/2305.01598}
\showURL{%
\tempurl}


\bibitem[Le and Gulwani(2014)]%
        {flashextract}
\bibfield{author}{\bibinfo{person}{Vu Le} {and} \bibinfo{person}{Sumit Gulwani}.} \bibinfo{year}{2014}\natexlab{}.
\newblock \showarticletitle{FlashExtract: a framework for data extraction by examples}. In \bibinfo{booktitle}{\emph{2014 Programming Language Design and Implementation}}. \bibinfo{publisher}{ACM}, \bibinfo{address}{New York, NY, USA}, \bibinfo{pages}{542--553}.
\newblock
\urldef\tempurl%
\url{https://www.microsoft.com/en-us/research/publication/flashextract-framework-data-extraction-examples/}
\showURL{%
\tempurl}


\bibitem[Li et~al\mbox{.}(2015)]%
        {vldb-query}
\bibfield{author}{\bibinfo{person}{Hao Li}, \bibinfo{person}{Chee-Yong Chan}, {and} \bibinfo{person}{David Maier}.} \bibinfo{year}{2015}\natexlab{}.
\newblock \showarticletitle{Query from Examples: An Iterative, Data-Driven Approach to Query Construction}.
\newblock \bibinfo{journal}{\emph{Proc. VLDB Endow.}} \bibinfo{volume}{8}, \bibinfo{number}{13} (\bibinfo{date}{sep} \bibinfo{year}{2015}), \bibinfo{pages}{2158–2169}.
\newblock
\showISSN{2150-8097}
\href{https://doi.org/10.14778/2831360.2831369}{doi:\nolinkurl{10.14778/2831360.2831369}}


\bibitem[Li et~al\mbox{.}(2023)]%
        {starcoder}
\bibfield{author}{\bibinfo{person}{Raymond Li}, \bibinfo{person}{Loubna~Ben Allal}, \bibinfo{person}{Yangtian Zi}, \bibinfo{person}{Niklas Muennighoff}, \bibinfo{person}{Denis Kocetkov}, \bibinfo{person}{Chenghao Mou}, \bibinfo{person}{Marc Marone}, \bibinfo{person}{Christopher Akiki}, \bibinfo{person}{Jia Li}, \bibinfo{person}{Jenny Chim}, \bibinfo{person}{Qian Liu}, \bibinfo{person}{Evgenii Zheltonozhskii}, \bibinfo{person}{Terry~Yue Zhuo}, \bibinfo{person}{Thomas Wang}, \bibinfo{person}{Olivier Dehaene}, \bibinfo{person}{Mishig Davaadorj}, \bibinfo{person}{Joel Lamy-Poirier}, \bibinfo{person}{João Monteiro}, \bibinfo{person}{Oleh Shliazhko}, \bibinfo{person}{Nicolas Gontier}, \bibinfo{person}{Nicholas Meade}, \bibinfo{person}{Armel Zebaze}, \bibinfo{person}{Ming-Ho Yee}, \bibinfo{person}{Logesh~Kumar Umapathi}, \bibinfo{person}{Jian Zhu}, \bibinfo{person}{Benjamin Lipkin}, \bibinfo{person}{Muhtasham Oblokulov}, \bibinfo{person}{Zhiruo Wang}, \bibinfo{person}{Rudra Murthy}, \bibinfo{person}{Jason
  Stillerman}, \bibinfo{person}{Siva~Sankalp Patel}, \bibinfo{person}{Dmitry Abulkhanov}, \bibinfo{person}{Marco Zocca}, \bibinfo{person}{Manan Dey}, \bibinfo{person}{Zhihan Zhang}, \bibinfo{person}{Nour Fahmy}, \bibinfo{person}{Urvashi Bhattacharyya}, \bibinfo{person}{Wenhao Yu}, \bibinfo{person}{Swayam Singh}, \bibinfo{person}{Sasha Luccioni}, \bibinfo{person}{Paulo Villegas}, \bibinfo{person}{Maxim Kunakov}, \bibinfo{person}{Fedor Zhdanov}, \bibinfo{person}{Manuel Romero}, \bibinfo{person}{Tony Lee}, \bibinfo{person}{Nadav Timor}, \bibinfo{person}{Jennifer Ding}, \bibinfo{person}{Claire Schlesinger}, \bibinfo{person}{Hailey Schoelkopf}, \bibinfo{person}{Jan Ebert}, \bibinfo{person}{Tri Dao}, \bibinfo{person}{Mayank Mishra}, \bibinfo{person}{Alex Gu}, \bibinfo{person}{Jennifer Robinson}, \bibinfo{person}{Carolyn~Jane Anderson}, \bibinfo{person}{Brendan Dolan-Gavitt}, \bibinfo{person}{Danish Contractor}, \bibinfo{person}{Siva Reddy}, \bibinfo{person}{Daniel Fried}, \bibinfo{person}{Dzmitry Bahdanau},
  \bibinfo{person}{Yacine Jernite}, \bibinfo{person}{Carlos~Muñoz Ferrandis}, \bibinfo{person}{Sean Hughes}, \bibinfo{person}{Thomas Wolf}, \bibinfo{person}{Arjun Guha}, \bibinfo{person}{Leandro von Werra}, {and} \bibinfo{person}{Harm de Vries}.} \bibinfo{year}{2023}\natexlab{}.
\newblock \bibinfo{title}{StarCoder: may the source be with you!}
\newblock
\showeprint[arxiv]{2305.06161}~[cs.CL]


\bibitem[Liguori et~al\mbox{.}(2022)]%
        {shellCodes}
\bibfield{author}{\bibinfo{person}{Pietro Liguori}, \bibinfo{person}{Erfan Al-Hossami}, \bibinfo{person}{Domenico Cotroneo}, \bibinfo{person}{Roberto Natella}, \bibinfo{person}{Bojan Cukic}, {and} \bibinfo{person}{Samira Shaikh}.} \bibinfo{year}{2022}\natexlab{}.
\newblock \showarticletitle{Can we generate shellcodes via natural language? An empirical study}.
\newblock \bibinfo{journal}{\emph{Automated Software Engineering}}  \bibinfo{volume}{29} (\bibinfo{year}{2022}), \bibinfo{pages}{1--34}.
\newblock


\bibitem[Lin(2006)]%
        {rel_formatting_1}
\bibfield{author}{\bibinfo{person}{Xiaofan Lin}.} \bibinfo{year}{2006}\natexlab{}.
\newblock \showarticletitle{Active layout engine: Algorithms and applications in variable data printing}.
\newblock \bibinfo{journal}{\emph{Computer-Aided Design}} \bibinfo{volume}{38}, \bibinfo{number}{5} (\bibinfo{year}{2006}), \bibinfo{pages}{444--456}.
\newblock


\bibitem[Luo et~al\mbox{.}(2023)]%
        {wizardcoder}
\bibfield{author}{\bibinfo{person}{Ziyang Luo}, \bibinfo{person}{Can Xu}, \bibinfo{person}{Pu Zhao}, \bibinfo{person}{Qingfeng Sun}, \bibinfo{person}{Xiubo Geng}, \bibinfo{person}{Wenxiang Hu}, \bibinfo{person}{Chongyang Tao}, \bibinfo{person}{Jing Ma}, \bibinfo{person}{Qingwei Lin}, {and} \bibinfo{person}{Daxin Jiang}.} \bibinfo{year}{2023}\natexlab{}.
\newblock \bibinfo{title}{WizardCoder: Empowering Code Large Language Models with Evol-Instruct}.
\newblock
\showeprint[arxiv]{2306.08568}~[cs.CL]


\bibitem[MacQueen et~al\mbox{.}(1967)]%
        {symbolic_clustering}
\bibfield{author}{\bibinfo{person}{James MacQueen} {et~al\mbox{.}}} \bibinfo{year}{1967}\natexlab{}.
\newblock \showarticletitle{Some methods for classification and analysis of multivariate observations}. In \bibinfo{booktitle}{\emph{Proceedings of the fifth Berkeley symposium on mathematical statistics and probability}}, Vol.~\bibinfo{volume}{1}. Oakland, CA, USA, \bibinfo{pages}{281--297}.
\newblock


\bibitem[Maddigan and Susnjak(2023)]%
        {table-viz}
\bibfield{author}{\bibinfo{person}{Paula Maddigan} {and} \bibinfo{person}{Teo Susnjak}.} \bibinfo{year}{2023}\natexlab{}.
\newblock \bibinfo{title}{Chat2VIS: Generating Data Visualisations via Natural Language using ChatGPT, Codex and GPT-3 Large Language Models}.
\newblock
\showeprint[arxiv]{2302.02094}~[cs.HC]


\bibitem[Mottin et~al\mbox{.}(2016)]%
        {vldb-search}
\bibfield{author}{\bibinfo{person}{Davide Mottin}, \bibinfo{person}{Matteo Lissandrini}, \bibinfo{person}{Yannis Velegrakis}, {and} \bibinfo{person}{Themis Palpanas}.} \bibinfo{year}{2016}\natexlab{}.
\newblock \showarticletitle{Exemplar Queries: A New Way of Searching}.
\newblock \bibinfo{journal}{\emph{The VLDB Journal}} \bibinfo{volume}{25}, \bibinfo{number}{6} (\bibinfo{date}{dec} \bibinfo{year}{2016}), \bibinfo{pages}{741–765}.
\newblock
\showISSN{1066-8888}
\href{https://doi.org/10.1007/s00778-016-0429-2}{doi:\nolinkurl{10.1007/s00778-016-0429-2}}


\bibitem[N.(2022)]%
        {spreadsheet-usage}
\bibfield{author}{\bibinfo{person}{Joseph N.}} \bibinfo{year}{2022}\natexlab{}.
\newblock \bibinfo{title}{Number of Google Sheets and Excel Users Worldwide}.
\newblock \bibinfo{howpublished}{\url{https://askwonder.com/research/number-google-s heets-users-worldwide-eoskdoxav}}.
\newblock
\newblock
\shownote{Last Accessed: 2022-07-30}.


\bibitem[Natarajan et~al\mbox{.}(2019)]%
        {ranking_outputs}
\bibfield{author}{\bibinfo{person}{Nagarajan Natarajan}, \bibinfo{person}{Danny Simmons}, \bibinfo{person}{Naren Datha}, \bibinfo{person}{Prateek Jain}, {and} \bibinfo{person}{Sumit Gulwani}.} \bibinfo{year}{2019}\natexlab{}.
\newblock \showarticletitle{Learning Natural Programs from a Few Examples in Real-Time}. In \bibinfo{booktitle}{\emph{AIStats}}. \bibinfo{publisher}{PMLR}, \bibinfo{address}{online}, \bibinfo{pages}{1714--1722}.
\newblock
\urldef\tempurl%
\url{https://www.microsoft.com/en-us/research/publication/learning-natural-programs-from-a-few-examples-in-real-time/}
\showURL{%
\tempurl}


\bibitem[Navigli et~al\mbox{.}(2023)]%
        {10.1145/3597307}
\bibfield{author}{\bibinfo{person}{Roberto Navigli}, \bibinfo{person}{Simone Conia}, {and} \bibinfo{person}{Bj\"{o}rn Ross}.} \bibinfo{year}{2023}\natexlab{}.
\newblock \showarticletitle{Biases in Large Language Models: Origins, Inventory, and Discussion}.
\newblock \bibinfo{journal}{\emph{J. Data and Information Quality}} \bibinfo{volume}{15}, \bibinfo{number}{2}, Article \bibinfo{articleno}{10} (\bibinfo{date}{jun} \bibinfo{year}{2023}), \bibinfo{numpages}{21}~pages.
\newblock
\showISSN{1936-1955}
\href{https://doi.org/10.1145/3597307}{doi:\nolinkurl{10.1145/3597307}}


\bibitem[Neuwirth and Arganbright(2003)]%
        {excelMathModel}
\bibfield{author}{\bibinfo{person}{Erich Neuwirth} {and} \bibinfo{person}{Deane Arganbright}.} \bibinfo{year}{2003}\natexlab{}.
\newblock \bibinfo{booktitle}{\emph{The Active Modeler: Mathematical Modeling With Microsoft Excel}}.
\newblock \bibinfo{publisher}{Duxbury Press}, \bibinfo{address}{online}.
\newblock


\bibitem[Nijkamp et~al\mbox{.}(2023)]%
        {codegen}
\bibfield{author}{\bibinfo{person}{Erik Nijkamp}, \bibinfo{person}{Bo Pang}, \bibinfo{person}{Hiroaki Hayashi}, \bibinfo{person}{Lifu Tu}, \bibinfo{person}{Huan Wang}, \bibinfo{person}{Yingbo Zhou}, \bibinfo{person}{Silvio Savarese}, {and} \bibinfo{person}{Caiming Xiong}.} \bibinfo{year}{2023}\natexlab{}.
\newblock \bibinfo{title}{CodeGen: An Open Large Language Model for Code with Multi-Turn Program Synthesis}.
\newblock
\showeprint[arxiv]{2203.13474}~[cs.LG]


\bibitem[OpenAI(2024)]%
        {openai2024gpt4}
\bibfield{author}{\bibinfo{person}{OpenAI}.} \bibinfo{year}{2024}\natexlab{}.
\newblock \bibinfo{title}{GPT-4 Technical Report}.
\newblock
\showeprint[arxiv]{2303.08774}~[cs.CL]


\bibitem[Pakman et~al\mbox{.}(2020)]%
        {neural_clustering}
\bibfield{author}{\bibinfo{person}{Ari Pakman}, \bibinfo{person}{Yueqi Wang}, \bibinfo{person}{Catalin Mitelut}, \bibinfo{person}{JinHyung Lee}, {and} \bibinfo{person}{Liam Paninski}.} \bibinfo{year}{2020}\natexlab{}.
\newblock \bibinfo{title}{Neural Clustering Processes}.
\newblock
\showeprint[arxiv]{1901.00409}~[stat.ML]


\bibitem[Payan et~al\mbox{.}(2023)]%
        {payan2023instructexcelbenchmarknaturallanguage}
\bibfield{author}{\bibinfo{person}{Justin Payan}, \bibinfo{person}{Swaroop Mishra}, \bibinfo{person}{Mukul Singh}, \bibinfo{person}{Carina Negreanu}, \bibinfo{person}{Christian Poelitz}, \bibinfo{person}{Chitta Baral}, \bibinfo{person}{Subhro Roy}, \bibinfo{person}{Rasika Chakravarthy}, \bibinfo{person}{Benjamin~Van Durme}, {and} \bibinfo{person}{Elnaz Nouri}.} \bibinfo{year}{2023}\natexlab{}.
\newblock \bibinfo{title}{InstructExcel: A Benchmark for Natural Language Instruction in Excel}.
\newblock
\showeprint[arxiv]{2310.14495}~[cs.CL]
\urldef\tempurl%
\url{https://arxiv.org/abs/2310.14495}
\showURL{%
\tempurl}


\bibitem[Poesia et~al\mbox{.}(2022)]%
        {synchromesh}
\bibfield{author}{\bibinfo{person}{Gabriel Poesia}, \bibinfo{person}{Oleksandr Polozov}, \bibinfo{person}{Vu Le}, \bibinfo{person}{Ashish Tiwari}, \bibinfo{person}{Gustavo Soares}, \bibinfo{person}{Christopher Meek}, {and} \bibinfo{person}{Sumit Gulwani}.} \bibinfo{year}{2022}\natexlab{}.
\newblock \showarticletitle{Synchromesh: Reliable code generation from pre-trained language models}.
\newblock \bibinfo{journal}{\emph{CoRR}}  \bibinfo{volume}{abs/2201.11227} (\bibinfo{year}{2022}).
\newblock
\showeprint[arXiv]{2201.11227}
\urldef\tempurl%
\url{https://arxiv.org/abs/2201.11227}
\showURL{%
\tempurl}


\bibitem[Rozière et~al\mbox{.}(2024)]%
        {codellama}
\bibfield{author}{\bibinfo{person}{Baptiste Rozière}, \bibinfo{person}{Jonas Gehring}, \bibinfo{person}{Fabian Gloeckle}, \bibinfo{person}{Sten Sootla}, \bibinfo{person}{Itai Gat}, \bibinfo{person}{Xiaoqing~Ellen Tan}, \bibinfo{person}{Yossi Adi}, \bibinfo{person}{Jingyu Liu}, \bibinfo{person}{Romain Sauvestre}, \bibinfo{person}{Tal Remez}, \bibinfo{person}{Jérémy Rapin}, \bibinfo{person}{Artyom Kozhevnikov}, \bibinfo{person}{Ivan Evtimov}, \bibinfo{person}{Joanna Bitton}, \bibinfo{person}{Manish Bhatt}, \bibinfo{person}{Cristian~Canton Ferrer}, \bibinfo{person}{Aaron Grattafiori}, \bibinfo{person}{Wenhan Xiong}, \bibinfo{person}{Alexandre Défossez}, \bibinfo{person}{Jade Copet}, \bibinfo{person}{Faisal Azhar}, \bibinfo{person}{Hugo Touvron}, \bibinfo{person}{Louis Martin}, \bibinfo{person}{Nicolas Usunier}, \bibinfo{person}{Thomas Scialom}, {and} \bibinfo{person}{Gabriel Synnaeve}.} \bibinfo{year}{2024}\natexlab{}.
\newblock \bibinfo{title}{Code Llama: Open Foundation Models for Code}.
\newblock
\showeprint[arxiv]{2308.12950}~[cs.CL]


\bibitem[Singh et~al\mbox{.}(2023c)]%
        {format5}
\bibfield{author}{\bibinfo{person}{Mukul Singh}, \bibinfo{person}{Jos{\'{e}} Cambronero}, \bibinfo{person}{Sumit Gulwani}, \bibinfo{person}{Vu Le}, \bibinfo{person}{Carina Negreanu}, \bibinfo{person}{Elnaz Nouri}, \bibinfo{person}{Mohammad Raza}, {and} \bibinfo{person}{Gust Verbruggen}.} \bibinfo{year}{2023}\natexlab{c}.
\newblock \showarticletitle{FormaT5: Abstention and Examples for Conditional Table Formatting with Natural Language}.
\newblock \bibinfo{journal}{\emph{Proc. {VLDB} Endow.}} \bibinfo{volume}{17}, \bibinfo{number}{3} (\bibinfo{year}{2023}), \bibinfo{pages}{497--510}.
\newblock
\urldef\tempurl%
\url{https://www.vldb.org/pvldb/vol17/p497-singh.pdf}
\showURL{%
\tempurl}


\bibitem[Singh et~al\mbox{.}(2023a)]%
        {codefusion}
\bibfield{author}{\bibinfo{person}{Mukul Singh}, \bibinfo{person}{Jos{\'e} Cambronero}, \bibinfo{person}{Sumit Gulwani}, \bibinfo{person}{Vu Le}, \bibinfo{person}{Carina Negreanu}, {and} \bibinfo{person}{Gust Verbruggen}.} \bibinfo{year}{2023}\natexlab{a}.
\newblock \showarticletitle{CodeFusion: A Pre-trained Diffusion Model for Code Generation}. In \bibinfo{booktitle}{\emph{Proceedings of the 2023 Conference on Empirical Methods in Natural Language Processing}}, \bibfield{editor}{\bibinfo{person}{Houda Bouamor}, \bibinfo{person}{Juan Pino}, {and} \bibinfo{person}{Kalika Bali}} (Eds.). \bibinfo{publisher}{Association for Computational Linguistics}, \bibinfo{address}{Singapore}, \bibinfo{pages}{11697--11708}.
\newblock
\href{https://doi.org/10.18653/v1/2023.emnlp-main.716}{doi:\nolinkurl{10.18653/v1/2023.emnlp-main.716}}


\bibitem[Singh et~al\mbox{.}(2023b)]%
        {datavinci}
\bibfield{author}{\bibinfo{person}{Mukul Singh}, \bibinfo{person}{José Cambronero}, \bibinfo{person}{Sumit Gulwani}, \bibinfo{person}{Vu Le}, \bibinfo{person}{Carina Negreanu}, {and} \bibinfo{person}{Gust Verbruggen}.} \bibinfo{year}{2023}\natexlab{b}.
\newblock \bibinfo{title}{DataVinci: Learning Syntactic and Semantic String Repairs}.
\newblock
\showeprint[arxiv]{2308.10922}~[cs.DB]


\bibitem[Singh et~al\mbox{.}(2022)]%
        {vehicle-tele}
\bibfield{author}{\bibinfo{person}{Mukul Singh}, \bibinfo{person}{Rahul~Kumar Dubey}, {and} \bibinfo{person}{Swarup Kumar}.} \bibinfo{year}{2022}\natexlab{}.
\newblock \showarticletitle{Chapter 15 - Vehicle telematics: An Internet of Things and Big Data approach}.
\newblock In \bibinfo{booktitle}{\emph{Artificial Intelligence and Machine Learning for EDGE Computing}}, \bibfield{editor}{\bibinfo{person}{Rajiv Pandey}, \bibinfo{person}{Sunil~Kumar Khatri}, \bibinfo{person}{Neeraj kumar Singh}, {and} \bibinfo{person}{Parul Verma}} (Eds.). \bibinfo{publisher}{Academic Press}, \bibinfo{pages}{235--254}.
\newblock
\showISBNx{978-0-12-824054-0}
\href{https://doi.org/10.1016/B978-0-12-824054-0.00019-8}{doi:\nolinkurl{10.1016/B978-0-12-824054-0.00019-8}}


\bibitem[Singh et~al\mbox{.}(2023e)]%
        {cornet}
\bibfield{author}{\bibinfo{person}{Mukul Singh}, \bibinfo{person}{Jos\'{e}~Cambronero S\'{a}nchez}, \bibinfo{person}{Sumit Gulwani}, \bibinfo{person}{Vu Le}, \bibinfo{person}{Carina Negreanu}, \bibinfo{person}{Mohammad Raza}, {and} \bibinfo{person}{Gust Verbruggen}.} \bibinfo{year}{2023}\natexlab{e}.
\newblock \showarticletitle{Cornet: Learning Table Formatting Rules By Example}.
\newblock \bibinfo{journal}{\emph{Proc. VLDB Endow.}} \bibinfo{volume}{16}, \bibinfo{number}{10} (\bibinfo{date}{jun} \bibinfo{year}{2023}), \bibinfo{pages}{2632–2644}.
\newblock
\showISSN{2150-8097}
\href{https://doi.org/10.14778/3603581.3603600}{doi:\nolinkurl{10.14778/3603581.3603600}}


\bibitem[Singh et~al\mbox{.}(2023d)]%
        {cornet-demo}
\bibfield{author}{\bibinfo{person}{Mukul Singh}, \bibinfo{person}{Jos\'{e}~Cambronero Sanchez}, \bibinfo{person}{Sumit Gulwani}, \bibinfo{person}{Vu Le}, \bibinfo{person}{Carina Negreanu}, {and} \bibinfo{person}{Gust Verbruggen}.} \bibinfo{year}{2023}\natexlab{d}.
\newblock \showarticletitle{Cornet: Learning Spreadsheet Formatting Rules by Example}.
\newblock \bibinfo{journal}{\emph{Proc. VLDB Endow.}} \bibinfo{volume}{16}, \bibinfo{number}{12} (\bibinfo{date}{aug} \bibinfo{year}{2023}), \bibinfo{pages}{4058–4061}.
\newblock
\showISSN{2150-8097}
\href{https://doi.org/10.14778/3611540.3611620}{doi:\nolinkurl{10.14778/3611540.3611620}}


\bibitem[Singha et~al\mbox{.}(2024a)]%
        {aligned-code-gen}
\bibfield{author}{\bibinfo{person}{Ananya Singha}, \bibinfo{person}{Bhavya Chopra}, \bibinfo{person}{Anirudh Khatry}, \bibinfo{person}{Sumit Gulwani}, \bibinfo{person}{Austin Henley}, \bibinfo{person}{Vu Le}, \bibinfo{person}{Chris Parnin}, \bibinfo{person}{Mukul Singh}, {and} \bibinfo{person}{Gust Verbruggen}.} \bibinfo{year}{2024}\natexlab{a}.
\newblock \showarticletitle{Semantically Aligned Question and Code Generation for Automated Insight Generation}. In \bibinfo{booktitle}{\emph{Proceedings of the 1st International Workshop on Large Language Models for Code}} (Lisbon, Portugal) \emph{(\bibinfo{series}{LLM4Code '24})}. \bibinfo{publisher}{Association for Computing Machinery}, \bibinfo{address}{New York, NY, USA}, \bibinfo{pages}{127–134}.
\newblock
\showISBNx{9798400705793}
\href{https://doi.org/10.1145/3643795.3648381}{doi:\nolinkurl{10.1145/3643795.3648381}}


\bibitem[Singha et~al\mbox{.}(2024b)]%
        {table-insights}
\bibfield{author}{\bibinfo{person}{Ananya Singha}, \bibinfo{person}{Bhavya Chopra}, \bibinfo{person}{Anirudh Khatry}, \bibinfo{person}{Sumit Gulwani}, \bibinfo{person}{Austin Henley}, \bibinfo{person}{Vu Le}, \bibinfo{person}{Chris Parnin}, \bibinfo{person}{Mukul Singh}, {and} \bibinfo{person}{Gust Verbruggen}.} \bibinfo{year}{2024}\natexlab{b}.
\newblock \showarticletitle{Semantically Aligned Question and Code Generation for Automated Insight Generation}. In \bibinfo{booktitle}{\emph{LLM4Code Workshop at ICSE '24}}.
\newblock
\urldef\tempurl%
\url{https://www.microsoft.com/en-us/research/publication/semantically-aligned-question-and-code-generation/}
\showURL{%
\tempurl}


\bibitem[Sun et~al\mbox{.}(2021)]%
        {TabNet}
\bibfield{author}{\bibinfo{person}{Kexuan Sun}, \bibinfo{person}{Harsha Rayudu}, {and} \bibinfo{person}{Jay Pujara}.} \bibinfo{year}{2021}\natexlab{}.
\newblock \showarticletitle{A Hybrid Probabilistic Approach for Table Understanding}.
\newblock \bibinfo{journal}{\emph{Proceedings of the AAAI Conference on Artificial Intelligence}} \bibinfo{volume}{35}, \bibinfo{number}{5} (\bibinfo{date}{May} \bibinfo{year}{2021}), \bibinfo{pages}{4366--4374}.
\newblock
\urldef\tempurl%
\url{https://ojs.aaai.org/index.php/AAAI/article/view/16562}
\showURL{%
\tempurl}


\bibitem[Sun et~al\mbox{.}(2023)]%
        {transcoder}
\bibfield{author}{\bibinfo{person}{Qiushi Sun}, \bibinfo{person}{Nuo Chen}, \bibinfo{person}{Jianing Wang}, \bibinfo{person}{Xiang Li}, {and} \bibinfo{person}{Ming Gao}.} \bibinfo{year}{2023}\natexlab{}.
\newblock \bibinfo{title}{TransCoder: Towards Unified Transferable Code Representation Learning Inspired by Human Skills}.
\newblock
\showeprint[arxiv]{2306.07285}~[cs.SE]


\bibitem[W3C(2020)]%
        {html-web-colors}
\bibfield{author}{\bibinfo{person}{W3C}.} \bibinfo{year}{2020}\natexlab{}.
\newblock \bibinfo{title}{CSS Color Module Level 3}.
\newblock
\urldef\tempurl%
\url{https://www.w3.org/TR/css-color-3/}
\showURL{%
\tempurl}


\bibitem[Wang et~al\mbox{.}(2023)]%
        {codet5+}
\bibfield{author}{\bibinfo{person}{Yue Wang}, \bibinfo{person}{Hung Le}, \bibinfo{person}{Akhilesh~Deepak Gotmare}, \bibinfo{person}{Nghi D.~Q. Bui}, \bibinfo{person}{Junnan Li}, {and} \bibinfo{person}{Steven C.~H. Hoi}.} \bibinfo{year}{2023}\natexlab{}.
\newblock \bibinfo{title}{CodeT5+: Open Code Large Language Models for Code Understanding and Generation}.
\newblock
\showeprint[arxiv]{2305.07922}~[cs.CL]


\bibitem[Wang et~al\mbox{.}(2021)]%
        {TUTA}
\bibfield{author}{\bibinfo{person}{Zhiruo Wang}, \bibinfo{person}{Haoyu Dong}, \bibinfo{person}{Ran Jia}, \bibinfo{person}{Jia Li}, \bibinfo{person}{Zhiyi Fu}, \bibinfo{person}{Shi Han}, {and} \bibinfo{person}{Dongmei Zhang}.} \bibinfo{year}{2021}\natexlab{}.
\newblock \showarticletitle{TUTA: Tree-Based Transformers for Generally Structured Table Pre-Training}. In \bibinfo{booktitle}{\emph{Proceedings of the 27th ACM SIGKDD Conference on Knowledge Discovery \&amp; Data Mining}} \emph{(\bibinfo{series}{KDD '21})}. \bibinfo{publisher}{Association for Computing Machinery}, \bibinfo{address}{New York, USA}, \bibinfo{pages}{1780–1790}.
\newblock
\showISBNx{9781450383325}
\href{https://doi.org/10.1145/3447548.3467434}{doi:\nolinkurl{10.1145/3447548.3467434}}


\bibitem[Wei et~al\mbox{.}(2022)]%
        {chain-of-thought}
\bibfield{author}{\bibinfo{person}{Jason Wei}, \bibinfo{person}{Xuezhi Wang}, \bibinfo{person}{Dale Schuurmans}, \bibinfo{person}{Maarten Bosma}, \bibinfo{person}{brian ichter}, \bibinfo{person}{Fei Xia}, \bibinfo{person}{Ed~H. Chi}, \bibinfo{person}{Quoc~V Le}, {and} \bibinfo{person}{Denny Zhou}.} \bibinfo{year}{2022}\natexlab{}.
\newblock \showarticletitle{Chain of Thought Prompting Elicits Reasoning in Large Language Models}. In \bibinfo{booktitle}{\emph{Advances in Neural Information Processing Systems}}, \bibfield{editor}{\bibinfo{person}{Alice~H. Oh}, \bibinfo{person}{Alekh Agarwal}, \bibinfo{person}{Danielle Belgrave}, {and} \bibinfo{person}{Kyunghyun Cho}} (Eds.).
\newblock
\urldef\tempurl%
\url{https://openreview.net/forum?id=_VjQlMeSB_J}
\showURL{%
\tempurl}


\bibitem[Yan and He(2020)]%
        {auto-suggest}
\bibfield{author}{\bibinfo{person}{Cong Yan} {and} \bibinfo{person}{Yeye He}.} \bibinfo{year}{2020}\natexlab{}.
\newblock \showarticletitle{Auto-Suggest: Learning-to-Recommend Data Preparation Steps Using Data Science Notebooks}. In \bibinfo{booktitle}{\emph{International Conference on Management of Data (SIGMOD)}}. \bibinfo{publisher}{ACM}, \bibinfo{pages}{1539--1554}.
\newblock
\urldef\tempurl%
\url{https://www.microsoft.com/en-us/research/publication/auto-suggest-learning-to-recommend-data-preparation-steps-using-data-science-notebooks/}
\showURL{%
\tempurl}


\bibitem[Yao et~al\mbox{.}(2023)]%
        {tree-of-thought}
\bibfield{author}{\bibinfo{person}{Shunyu Yao}, \bibinfo{person}{Dian Yu}, \bibinfo{person}{Jeffrey Zhao}, \bibinfo{person}{Izhak Shafran}, \bibinfo{person}{Thomas~L. Griffiths}, \bibinfo{person}{Yuan Cao}, {and} \bibinfo{person}{Karthik Narasimhan}.} \bibinfo{year}{2023}\natexlab{}.
\newblock \bibinfo{title}{Tree of Thoughts: Deliberate Problem Solving with Large Language Models}.
\newblock
\showeprint[arxiv]{2305.10601}~[cs.CL]


\bibitem[Yin et~al\mbox{.}(2020)]%
        {TaBERT}
\bibfield{author}{\bibinfo{person}{Pengcheng Yin}, \bibinfo{person}{Graham Neubig}, \bibinfo{person}{Wen-tau Yih}, {and} \bibinfo{person}{Sebastian Riedel}.} \bibinfo{year}{2020}\natexlab{}.
\newblock \showarticletitle{{T}a{BERT}: Pretraining for Joint Understanding of Textual and Tabular Data}. In \bibinfo{booktitle}{\emph{Proceedings of the 58th Annual Meeting of the Association for Computational Linguistics}}. \bibinfo{publisher}{Association for Computational Linguistics}, \bibinfo{address}{Online}, \bibinfo{pages}{8413--8426}.
\newblock
\href{https://doi.org/10.18653/v1/2020.acl-main.745}{doi:\nolinkurl{10.18653/v1/2020.acl-main.745}}


\bibitem[Zhao et~al\mbox{.}(2024)]%
        {nl2formula}
\bibfield{author}{\bibinfo{person}{Wei Zhao}, \bibinfo{person}{Zhitao Hou}, \bibinfo{person}{Siyuan Wu}, \bibinfo{person}{Yan Gao}, \bibinfo{person}{Haoyu Dong}, \bibinfo{person}{Yao Wan}, \bibinfo{person}{Hongyu Zhang}, \bibinfo{person}{Yulei Sui}, {and} \bibinfo{person}{Haidong Zhang}.} \bibinfo{year}{2024}\natexlab{}.
\newblock \bibinfo{title}{NL2Formula: Generating Spreadsheet Formulas from Natural Language Queries}.
\newblock
\showeprint[arxiv]{2402.14853}~[cs.CL]
\urldef\tempurl%
\url{https://arxiv.org/abs/2402.14853}
\showURL{%
\tempurl}


\end{thebibliography}

\end{document}